\def\fiat{}
\fiat
\def\ap{{\it a priori\ }}

\let\st=\scriptstyle
\font\tenmib=cmmib10
\font\sevenmib=cmmib10 scaled 800
\font\titolo=cmbx12
\font\titolone=cmbx10 scaled\magstep 2
\font\cs=cmcsc10
\font\css=cmcsc8

\font\ninerm=cmr9
\font\ottorm=cmr8
\textfont5=\tenmib
\scriptfont5=\sevenmib
\scriptscriptfont5=\fivei

\font\msytw=msbm10 scaled\magstep1

\font\msytwww=msbm7 scaled\magstep1
\font\indbf=cmbx10 scaled\magstep2

\newskip\ttglue
\font\ottorm=cmr8\font\ottoi=cmmi7\font\ottosy=cmsy7
\font\ottobf=cmbx7\font\ottott=cmtt8\font\ottosl=cmsl8\font\ottoit=cmti7
\font\sixrm=cmr6\font\sixbf=cmbx7\font\sixi=cmmi7\font\sixsy=cmsy7
\font\fiverm=cmr5\font\fivesy=cmsy5\font\fivei=cmmi5\font\fivebf=cmbx5

\def\ottopunti{\def\rm{\fam0\ottorm}\textfont0=\ottorm%
\scriptfont0=\sixrm\scriptscriptfont0=\fiverm\textfont1=\ottoi%
\scriptfont1=\sixi\scriptscriptfont1=\fivei\textfont2=\ottosy%
\scriptfont2=\sixsy\scriptscriptfont2=\fivesy\textfont3=\tenex%
\scriptfont3=\tenex\scriptscriptfont3=\tenex\textfont\itfam=\ottoit%
\def\it{\fam\itfam\ottoit}\textfont\slfam=\ottosl%
\def\sl{\fam\slfam\ottosl}\textfont\ttfam=\ottott%
\def\tt{\fam\ttfam\ottott}\textfont\bffam=\ottobf%
\scriptfont\bffam=\sixbf\scriptscriptfont\bffam=\fivebf%
\def\bf{\fam\bffam\ottobf}\tt\ttglue=.5em plus.25em minus.15em%
\setbox\strutbox=\hbox{\vrule height7pt height2pt width0pt}%
\normalbaselineskip=9pt\let\sc=\sixrm\normalbaselines\rm}
\let\nota=\ottopunti

%
%
%
%
%
%
%

\global\newcount\numsec\global\newcount\numapp
\global\newcount\numfor\global\newcount\numfig\global\newcount\numsub
\numsec=0\numapp=0\numfig=1
\def\veroparagrafo{\number\numsec}\def\veraformula{\number\numfor}
\def\veraappendice{\number\numapp}\def\verasub{\number\numsub}
\def\verafigura{\number\numfig}

\def\section(#1,#2){\advance\numsec by 1\numfor=1\numsub=1%
\SIA p,#1,{\veroparagrafo} %
\write15{\string\Fp (#1){\secc(#1)}}%
\write16{ sec. #1 ==> \secc(#1)  }%
\hbox to \hsize{\titolo\hfill \number\numsec. #2\hfill%
\expandafter{\alato(sec. #1)}}\*}

\def\appendix(#1,#2){\advance\numapp by 1\numfor=1\numsub=1%
\SIA p,#1,{A\veraappendice} %
\write15{\string\Fp (#1){\secc(#1)}}%
\write16{ app. #1 ==> \secc(#1)  }%
\hbox to \hsize{\titolo\hfill Appendix A\number\numapp. #2\hfill%
\expandafter{\alato(app. #1)}}\*}

\def\senondefinito#1{\expandafter\ifx\csname#1\endcsname\relax}

\def\SIA #1,#2,#3 {\senondefinito{#1#2}%
\expandafter\xdef\csname #1#2\endcsname{#3}\else
\write16{???? ma #1#2 e' gia' stato definito !!!!} \fi}

\def \Fe(#1)#2{\SIA fe,#1,#2 }
\def \Fp(#1)#2{\SIA fp,#1,#2 }
\def \Fg(#1)#2{\SIA fg,#1,#2 }

\def\etichetta(#1){(\veroparagrafo.\veraformula)%
\SIA e,#1,(\veroparagrafo.\veraformula) %
\global\advance\numfor by 1%
\write15{\string\Fe (#1){\equ(#1)}}%
\write16{ EQ #1 ==> \equ(#1)  }}

\def\etichettaa(#1){(A\veraappendice.\veraformula)%
\SIA e,#1,(A\veraappendice.\veraformula) %
\global\advance\numfor by 1%
\write15{\string\Fe (#1){\equ(#1)}}%
\write16{ EQ #1 ==> \equ(#1) }}

\def\getichetta(#1){Fig. \verafigura%
\SIA g,#1,{\verafigura} %
\global\advance\numfig by 1%
\write15{\string\Fg (#1){\graf(#1)}}%
\write16{ Fig. #1 ==> \graf(#1) }}

\def\etichettap(#1){\veroparagrafo.\verasub%
\SIA p,#1,{\veroparagrafo.\verasub} %
\global\advance\numsub by 1%
\write15{\string\Fp (#1){\secc(#1)}}%
\write16{ par #1 ==> \secc(#1)  }}

\def\etichettapa(#1){A\veraappendice.\verasub%
\SIA p,#1,{A\veraappendice.\verasub} %
\global\advance\numsub by 1%
\write15{\string\Fp (#1){\secc(#1)}}%
\write16{ par #1 ==> \secc(#1)  }}

\def\Eq(#1){\eqno{\etichetta(#1)\alato(#1)}}
\def\eq(#1){\etichetta(#1)\alato(#1)}
\def\Eqa(#1){\eqno{\etichettaa(#1)\alato(#1)}}
\def\eqa(#1){\etichettaa(#1)\alato(#1)}
\def\eqg(#1){\getichetta(#1)\alato(fig. #1)}
\def\sub(#1){\0\palato(p. #1){\bf \etichettap(#1).}}
\def\asub(#1){\0\palato(p. #1){\bf \etichettapa(#1).}}

\def\equv(#1){\senondefinito{fe#1}$\clubsuit$#1%
\write16{eq. #1 non e' (ancora) definita}%
\else\csname fe#1\endcsname\fi}
\def\grafv(#1){\senondefinito{fg#1}$\clubsuit$#1%
\write16{fig. #1 non e' (ancora) definito}%
\else\csname fg#1\endcsname\fi}
\def\secv(#1){\senondefinito{fp#1}$\clubsuit$#1%
\write16{par. #1 non e' (ancora) definito}%
\else\csname fp#1\endcsname\fi}

\def\equ(#1){\senondefinito{e#1}\equv(#1)\else\csname e#1\endcsname\fi}
\def\graf(#1){\senondefinito{g#1}\grafv(#1)\else\csname g#1\endcsname\fi}
\def\secc(#1){\senondefinito{p#1}\secv(#1)\else\csname p#1\endcsname\fi}
\def\sec(#1){{\S\secc(#1)}}

\def\BOZZA{\bz=1
\def\alato(##1){\rlap{\kern-\hsize\kern-1.2truecm{$\scriptstyle##1$}}}
\def\palato(##1){\rlap{\kern-1.2truecm{$\scriptstyle##1$}}}
}

\def\alato(#1){}
\def\galato(#1){}
\def\palato(#1){}


{\count255=\time\divide\count255 by 60 \xdef\hourmin{\number\count255}
        \multiply\count255 by-60\advance\count255 by\time
   \xdef\hourmin{\hourmin:\ifnum\count255<10 0\fi\the\count255}}

\def\oramin{\hourmin }

\def\data{\number\day/\ifcase\month\or gennaio \or febbraio \or marzo \or
aprile \or maggio \or giugno \or luglio \or agosto \or settembre
\or ottobre \or novembre \or dicembre \fi/\number\year;\ \oramin}
\setbox200\hbox{$\scriptscriptstyle \data $}
\footline={\rlap{\hbox{\copy200}}\tenrm\hss \number\pageno\hss}

%
%
%
\def\ins#1#2#3{\vbox to0pt{\kern-#2 \hbox{\kern#1 #3}\vss}\nointerlineskip}
%
%
%
\newdimen\xshift \newdimen\xwidth \newdimen\yshift

\def\insertplot#1#2#3#4#5{\par%
\xwidth=#1 \xshift=\hsize \advance\xshift by-\xwidth \divide\xshift by 2%
\yshift=#2 \divide\yshift by 2%
\line{\hskip\xshift \vbox to #2{\vfil%
#3 \includegraphics{#4.ps}}\hfill \raise\yshift\hbox{#5}}}

\def\eqfig#1#2#3#4#5{ \par\xwidth=#1
\xshift=\hsize \advance\xshift by-\xwidth \divide\xshift by 2
\yshift=#2 \divide\yshift by 2 \line{\hglue\xshift \vbox to #2{\vfil #3
\includegraphics{#4.ps} }\hfill\raise\yshift\hbox{#5}}} 

\def\8{\write13}


\let\a=\alpha \let\b=\beta  \let\g=\gamma     \let\d=\delta  \let\e=\varepsilon
  \let\h=\eta   \let\th=\theta \let\k=\kappa   \let\l=\lambda
    \let\n=\nu    \let\x=\xi        \let\p=\pi      
\let\s=\sigma \let\t=\tau   \let\f=\varphi     
   \let\o=\omega 
 \let\D=\Delta  \let\Th=\Theta   \let\L=\Lambda

\def\\{\hfill\break} \let\==\equiv

\let\io=\infty 

\let\0=\noindent \def\pagina{{\vfill\eject}}

\def\ie{\hbox{\it i.e.\ }}\def\eg{\hbox{\it e.g.\ }}
\let\dpr=\partial 

\def\tende#1{\,\vtop{\ialign{##\crcr\rightarrowfill\crcr
 \noalign{\kern-1pt\nointerlineskip}
 \hskip3.pt${\scriptstyle #1}$\hskip3.pt\crcr}}\,}
\def\otto{\,{\kern-1.truept\leftarrow\kern-5.truept\to\kern-1.truept}\,}
\def\fra#1#2{{#1\over#2}}

\def\PP{{\cal P}} \def\VV{{\cal V}}
\def\FF{{\cal F}}\def\HHH{{\cal H}}
\def\TT{{\cal T}}\def\BB{{\cal B}} 
\def\RR{{\cal R}}\def\LL{{\cal L}} 
\def\AAA{{\cal A}}\def\GG{{\cal G}} 

\def\T#1{{#1_{\kern-3pt\lower7pt\hbox{$\widetilde{}$}}\kern3pt}}
\def\VVV#1{{\underline #1}_{\kern-3pt
\lower7pt\hbox{$\widetilde{}$}}\kern3pt\,}
\def\W#1{#1_{\kern-3pt\lower7.5pt\hbox{$\widetilde{}$}}\kern2pt\,}
\def\Re{{\rm Re}\,}\def\Im{{\rm Im}\,}

\def\indica{\leaders \hbox to 0.5cm{\hss.\hss}\hfill}
\def\guida{\leaders\hbox to 1em{\hss.\hss}\hfill}

\def\V#1{{\bf #1}}
\def\aa{{\V \a}}\def\hh{{\V h}}\def\AA{{\V A}}
\def\GG{{\V G}}\def\BB{{\V B}}\def\aaa{{\V a}}\def\bbb{{\V b}}
\def\nn{{\V \n}}

\def\ohh{\ol{\hh}}
\def\Val{{\bf Val}}

\def\oV{{\ol{{\Val}}}}

\mathchardef\aa = "050B
\mathchardef\bb = "050C
\mathchardef\xxx= "0518
\mathchardef\zz = "0510
\mathchardef\oo = "0521
\mathchardef\ll = "0515
\mathchardef\mmm= "0516
\mathchardef\Dp = "0540
\mathchardef\H  = "0548
\mathchardef\FFF= "0546
\mathchardef\ppp= "0570
\mathchardef\nn = "0517
\mathchardef\pps= "0520
\mathchardef\XXX= "0504
\mathchardef\FFF= "0508
\mathchardef\tth= "0512

\def\ol#1{{\overline #1}}

\def\der{{\rm d}}

\let\ciao=\bye
\def\qed{\raise1pt\hbox{\vrule height5pt width5pt depth0pt}}

\def\indic{\hbox{\raise-2pt \hbox{\indbf 1}}}

\def\RRR{\hbox{\msytw R}}

\def\NNN{\hbox{\msytw N}} 
\def\nnn{\hbox{\msytwww N}} \def\ZZZ{\hbox{\msytw Z}}
 \def\zzz{\hbox{\msytwww Z}}
\def\TTT{\hbox{\msytw T}}

\newcount\mgnf  
\mgnf=0 

\ifnum\mgnf=0
\def\openone{\leavevmode\hbox{\ninerm 1\kern-3.3pt\tenrm1}}%
\def\*{\vglue0.3truecm}\fi
\ifnum\mgnf=1
\def\openone{\leavevmode\hbox{\ninerm 1\kern-3.63pt\tenrm1}}%
\def\*{\vglue0.5truecm}\fi

\newcount\tipobib\newcount\bz\bz=0\newcount\aux\aux=1
\newdimen\bibskip\newdimen\maxit\maxit=0pt


\tipobib=0
\def\9#1{\ifnum\aux=1#1\else\relax\fi}

\newwrite\bib
\immediate\openout\bib=\jobname.bib
\global\newcount\bibex
\bibex=0
\def\verabib{\number\bibex}

\ifnum\tipobib=0
\def\cita#1{\expandafter\ifx\csname c#1\endcsname\relax
\hbox{$\clubsuit$}#1\write16{Manca #1 !}%
\else\csname c#1\endcsname\fi}
\def\rife#1#2#3{\immediate\write\bib{\string\raf{#2}{#3}{#1}}
\immediate\write15{\string\C(#1){[#2]}}
\setbox199=\hbox{#2}\ifnum\maxit < \wd199 \maxit=\wd199\fi}
\else
\def\cita#1{%
\expandafter\ifx\csname d#1\endcsname\relax%
\expandafter\ifx\csname c#1\endcsname\relax%
\hbox{$\clubsuit$}#1\write16{Manca #1 !}%
\else\probib(ref. numero )(#1)%
\csname c#1\endcsname%
\fi\else\csname d#1\endcsname\fi}%
\def\rife#1#2#3{\immediate\write15{\string\Cp(#1){%
\string\immediate\string\write\string\bib{\string\string\string\raf%
{\string\verabib}{#3}{#1}}%
\string\Cn(#1){[\string\verabib]}%
\string\CCc(#1)%
}}}%
\fi

\def\Cn(#1)#2{\expandafter\xdef\csname d#1\endcsname{#2}}
\def\CCc(#1){\csname d#1\endcsname}
\def\probib(#1)(#2){\global\advance\bibex+1%
\9{\immediate\write16{#1\verabib => #2}}%
}

\def\C(#1)#2{\SIA c,#1,{#2}}
\def\Cp(#1)#2{\SIAnx c,#1,{#2}}

\def\SIAnx #1,#2,#3 {\senondefinito{#1#2}%
\expandafter\def\csname#1#2\endcsname{#3}\else%
\write16{???? ma #1,#2 e' gia' stato definito !!!!}\fi}

\bibskip=10truept
\def\hboxto{\hbox to}

\catcode`\{=12\catcode`\}=12
\catcode`\<=1\catcode`\>=2
\immediate\write\bib<
	\string\halign{\string\hboxto \string\maxit%
	{\string #\string\hfill}&%
        \string\vtop{\string\parindent=0pt\string\advance\string\hsize%
	by -1.55truecm%
	\string#\string\vskip \bibskip
	}\string\cr%
>
\catcode`\{=1\catcode`\}=2
\catcode`\<=12\catcode`\>=12

\def\raf#1#2#3{\ifnum \bz=0 [#1]&#2 \cr\else
\llap{${}_{\rm #3}$}[#1]&#2\cr\fi}

\newread\bibin

\catcode`\{=12\catcode`\}=12
\catcode`\<=1\catcode`\>=2
\def\chiudibib<
\catcode`\{=12\catcode`\}=12
\catcode`\<=1\catcode`\>=2
\immediate\write\bib<}>
\catcode`\{=1\catcode`\}=2
\catcode`\<=12\catcode`\>=12
>
\catcode`\{=1\catcode`\}=2
\catcode`\<=12\catcode`\>=12

\def\makebiblio{
\ifnum\tipobib=0
\advance \maxit by 10pt
\else
\maxit=1.truecm
\fi
\chiudibib
\immediate \closeout\bib
\openin\bibin=\jobname.bib
\ifeof\bibin\relax\else
\raggedbottom
\input \jobname.bib
\fi
}

\openin13=#1.aux \ifeof13 \relax \else
\input #1.aux \closein13\fi
\openin14=\jobname.aux \ifeof14 \relax \else
\input \jobname.aux \closein14 \fi
\immediate\openout15=\jobname.aux

\def\biblio{\*\*\centerline{\titolo References}\*\nobreak\makebiblio}


\ifnum\mgnf=0
   \magnification=\magstep0
   \hsize=15truecm\vsize=24.truecm
   \parindent=0.3cm\baselineskip=0.45cm\fi
\ifnum\mgnf=1
   \magnification=\magstep1\hoffset=0.truecm
   \hsize=15truecm\vsize=24.truecm
   \baselineskip=18truept plus0.1pt minus0.1pt \parindent=0.9truecm
   \lineskip=0.5truecm\lineskiplimit=0.1pt      \parskip=0.1pt plus1pt\fi

\centerline{\titolone Hyperbolic low-dimensional invariant tori}
\vskip.2truecm
\centerline{\titolone and summations of divergent series}

\vskip1.truecm \centerline{{\titolo
G. Gallavotti $^{*}$ and G. Gentile $^{\dagger}$}}
\vskip.2truecm
\centerline{{}$^{*}$
INFN, Dipartimento di Fisica,
Universit\`a di Roma 1,
I-00185}
\vskip.1truecm
\centerline{{}$^{\dagger}$
Dipartimento di Matematica,
Universit\`a di Roma 3,
I-00146}
\vskip1.truecm

\line{\vtop{
\line{\hskip1.5truecm\vbox{\advance \hsize by -3.1 truecm
\0{\cs Abstract.}
{\it We consider a class of {\sl a priori} stable quasi-integrable
analytic Hamiltonian systems and study the regularity of
low-dimensional hyperbolic invariant tori as functions of the
perturbation parameter.  We show that, under natural nonresonance
conditions, such tori exist and can be identified through the maxima
or minima of a suitable potential. They are analytic inside a disc
centered at the origin and deprived of a region around the positive or
negative real axis with a quadratic cusp at the origin. The invariant
tori admit an asymptotic series at the origin with Taylor coefficients
that grow at most as a power of a factorial and a remainder that to
any order $N$ is bounded by the $(N+1)$-st power of the argument
times a power of $N!$. We show the existence of a summation criterion
of the (generically divergent) series, in powers of the perturbation
size, that represent the parametric equations of the tori by following
the renormalization group methods for the resummations of perturbative
series in quantum field theory.}}
\hfill} }}

\vskip1.truecm

\section(1,Introduction)

\sub(1.1) {\it The model.}
Consider the Hamiltonian
$$ \HHH = \oo\cdot\AA + {1\over 2} \AA \cdot \AA +
{1\over 2} \BB\cdot\BB + \e f(\aa,\bb) ,
\Eq(1.1) $$
where $(\aa,\AA)\in \TTT^{r}\times\RRR^{r}$ and
$(\bb,\BB)\in\TTT^{s}\times\RRR^{s}$ are conjugated variables,
$\cdot$ denotes the inner product both in $\RRR^{r}$ and in $\RRR^{s}$,
and $\oo$ is a vector in $\RRR^{r}$ satisfying the Diophantine condition
$$ \left| \oo\cdot\nn \right| > C_{0} |\nn|^{-\t} \qquad
\forall \nn\in \ZZZ^{r}\setminus \{\V0\} ,
\Eq(1.2) $$
with $C_{0}>0$ and $\t\ge r-1$; we shall define by $D_{\t}(C_{0})$ the set
of rotation vectors in $\RRR^{r}$ satisfying \equ(1.2). We also write
$$ f(\aa,\bb) = \sum_{\nn\in\zzz^{r}} e^{i\nn\cdot\aa}
f_{\nn}(\bb) ,
\Eq(1.3) $$
and set $d=r+s$. We shall suppose that $f$ is analytic in a strip of
width $\k>0$ around the real axis of the variables $\aa,\bb$,
so that there exists a constant $F$ such that
$|f_{\nn}(\bb)|\le F\,e^{-\k|\nn|}$ for all $\nn\in\ZZZ^{r}$
and all $\bb\in\TTT^{s}$.

\*

\sub(1.2) {\it Low-dimensional tori.}
The equations of motion for the system \equ(1.1) are
$$ \cases{
\dot \aa = \oo + \AA , & \cr
\dot \bb = \BB , & \cr
\dot \AA = - \e\dpr_{\aa}f(\aa,\bb) , & \cr
\dot \BB = - \e\dpr_{\bb}f(\aa,\bb) , & \cr}
\Eq(1.4) $$
For $\e=0$ the system of equations \equ(1.4), with initial data
$(\aa_0,\bb_{0},\V0,\V0)$, admits the solution
$$ \cases{
\aa(t) = \aa_{0} + \oo t , & \cr
\bb(t) = \bb_{0} , & \cr
\AA(t) = \V0 , & \cr
\BB(t) = \V0 , & \cr}
\Eq(1.5) $$
which corresponds to a $r$-dimensional torus (KAM torus): the first
$r$ angles rotate with angular velocity $\o_{1},\ldots,\o_{r}$, while
the remaining $s$ remain fixed to their initial values.

Note that \equ(1.4) can be written as
$$ \cases{
\ddot \aa = - \e\dpr_{\aa} f(\aa,\bb) , & \cr 
\ddot \bb = - \e\dpr_{\bb} f(\aa,\bb) . & \cr}
\Eq(1.6) $$
so that we obtain closed equations for the angle variables:
once a solution has been found for them, it can be used
to find the action components by a simple integration.

We look for solutions of \equ(1.6), for $\e\neq 0$,
conjugated to \equ(1.5) , \ie
we look for solutions of the form
$$ \cases{
\aa(t) = \pps + \aaa (\pps,\bb_{0}; \e) , & \cr
\bb(t) = \bb_{0} + \bbb(\pps,\bb_{0}; \e) , & \cr}
\Eq(1.7) $$
for some functions $\aaa$ and $\bbb$, real analytic
and $2\p$-periodic in $\pps\in\TTT^{r}$, such that the motion
in the variable $\pps$ is $\dot\pps=\oo$.

We shall prove the following result.

\*

\sub(1.3) {\cs Theorem.}
{\it Consider the equations of motion \equ(1.6) for
$\oo\in D_{\t}(C_{0})$, and suppose $\bb_{0}$ to be such that
$$ \eqalign{
& \dpr_{\bb}f_{\V0}(\bb_{0}) = \V0 , \cr
& \dpr_{\bb}^{2} f_{\V0}(\bb_{0}) \hbox{ is negative definite.} \cr}
\Eq(1.8) $$
There exist a constant $\e_{0}>0$ and,
for all $\e\in(0,\e_{0})$, two functions
$\aaa(\pps,\bb_{0};\e)$ and $\bbb(\pps,\bb_{0};\e)$,
real analytic and $2\p$-periodic in $\pps\in\TTT^{r}$,
such that \equ(1.7) is a solution of \equ(1.6).}
\*

\sub(1.4) {\it Remarks.} (1) As it is well known, and as it 
will appear from the proof, the solutions whose existence is stated by
the theorem cannot be expected to be analytic in $\e$ at
$\e=0$. Furthermore, if the second condition in \equ(1.8) is replaced
with
$$ \dpr_{\bb}^{2} f_{\V0}(\bb_{0})\ \hbox{\it is positive definite, }
\Eq(1.9) $$
then the same conclusions hold for $\e\in(-\e_{0},0)$.
\\
(2) The proof will yield more detailed
information on the regularity of the considered tori, as we shall
point out.  In particular the analyticity domain is much larger, see
the heart-like domain $D_{0}$ in Figure 1 below
(and the discussion in the forthcoming Section \secc(5.9)).
In fact we think that our technique can lead to prove existence of
many elliptic invariant tori, \ie for a large set of negative $\e$'s,
and to understand some of their analyticity properties;
see Section \secc(6) for further remarks and results.

\midinsert
\*
\eqfig{140pt}{70pt}
{\ins{60pt}{60pt}{$ \rm complex \atop \e\rm-plane$}
\ins{18pt}{28pt}{$\st O$}
}{fig1}{}
\*\*
\line{\vtop{\line{\hskip2.5truecm\vbox{\advance\hsize by -5.1 truecm
\0{{\css Fig. 1.}
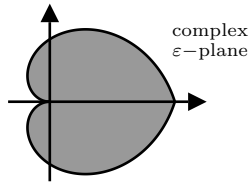
{\nota Analyticity domain $D_{0}$ for the hyperbolic invariant torus.
The cusp at the origin is a second order cusp. The figure corresponds
to the case in \equ(1.8) of the theorem \secc(1.3).\vfill}}
} \hfill} }}
\*
\endinsert

\*

\sub(1.5) {\it Contents of the paper.}
The paper is organized as follows. In Section \secc(2) we introduce
the main graph techniques which will be used, and we prove the
formal solubility of the equatins of motions: this is enough
if one wants to prove existence and analyticity of periodic solutions,
(see the remark \secc(2.10) below), but it requires
new arguments to obtain existence of quasi-periodic solutions.
Such arguments are developped in the following Sections:
in Section \secc(3) we introduce the concept of self-energy graph,
which will play a crucial r\^ole, and we describe
the basic cancellation mechanisms which will be used in
Section \secc(4) to perform a suitable resummation of the series.
In Section \secc(5) we shall use such results in order
to prove the convergence of the resummed series 
and its analyticity properties. Finally in Section \secc(6)
we make some conclusive remarks, and briefly discuss
possible generalizations and extensions of the results.
The main technical aspects of the proof will be relegated
to the Appendices.

\*

\sub(1.6) {\it Comparison with other papers.}
The problem considered here is {\it a priori stable} in the sense of
\cita{CG}. While the problem considered in the literature corresponds to
a Hamiltonian of the form
$$ \fra12 \AA^2+\oo\cdot\V A+ \fra12 \BB^{2} -\fra12 \L \bb^{2}
+O(\bb^{3}+\BB^{3}) ,
\Eq(1.10) $$
where $\L$ is a, {\it a priori fixed}, nondegenerate matrix (so that
the invariant torus at $\V A=\V0$ has \ap a well defined stability
property, \ie its elliptic or hyperbolic stability is already well
defined before the perturbation is switched on); this case
is called {\it a priori unstable} in \cita{CG}. The system
\equ(1.10) has been widely studied in the hyperbolic case $\L>0$, in
the elliptic case $\L<0$, and in the mixed case. The general hyperbolic
problem has been studied in \cita{Mo}; in \cita{Gr} the stable and
unstable manifolds of the tori are also determined. The elliptic and
mixed cases have been considered in great detail in several papers
starting with \cita{Me}; the reader will find, besides original
results, a complete description of the subsequent results and the
relevant references in the recent paper \cita{R}, with some very
recent further results, on a subject that keeps being under intense
study, in \cita{XY} and \cita{BKS}.

Our case can be reduced to the \equ(1.10) with $\L$ replaced by $\e\L$;
in our case the perturbation is small because it is proportional to
$\e$, while in \equ(1.10) one makes use of the possibility of
taking $\bb$ small to obtain a small perturbation.
This in turn can be reduced by
classical perturbation analysis to the theory of \equ(1.10).

We consider the novelty of this paper to be the technical analysis of
the analyticity, in $\e$, of the resonant (\ie of dimension lower than
maximal) hyperbolic invariant tori of \equ(1.1) in a region as
large as Figure 1 based on the Lindstedt series
method; the same analyticity domain can possibly be obtained by a
careful analysis of the proofs of the paper \cita{Mo}.  

In fact one is also interested in asking whether the analyticity
region in $\e$ can be extended further to reach {\it some} points on
the negative real axis and whether the analytic continuation to $\e<0$
of the parametric equations of the hyperbolic tori can be interpreted
as the parametric equations of elliptic tori. We do not address the
latter question: the analysis performed in the present paper at first
suggests us the following conjecture.

\*

\sub(1.7) {\cs Conjecture.}
{\it Consider the equations of motion \equ(1.6) for $\oo\in
D_{\t}(C_{0})$, and suppose $\bb_{0}$ to be such that
$$ \eqalign{
& \dpr_{\bb}f_{\V0}(\bb_{0}) = \V0 , \cr
& \dpr_{\bb}^{2} f_{\V0}(\bb_{0}) \hbox{ is negative definite.} \cr}
\Eq(1.11) $$
Is it possible that there exist constants $\e_{0}>0, \x>0$ and a
subset $I_{\e_0} \subset [-\e_0,0]$ with length $\ge
(1-\e_0^\x)\,\e_0$ such that the functions of the theorem \secc(1.3)
above are analytically continuable outside the domain $D_{0}$ along
vertical lines which start at points interior to $D_{0}$ and end on
$I_{\e_0}$, where their boundary value is real and gives the
parametric equations of an invariant torus for all $\e\in I_{\e_0}$ on
which the motion according to \equ(1.6) is $\dot\pps=\oo$?}

\*

The extended domain shape, near the origin, suggested
in the above conjecture is illustrated in the following Figure 1$'$.
\pagina

\midinsert
\*
\eqfig{95pt}{85pt}{
\ins{75pt}{95pt}{$ \rm complex \atop \e\rm-plane$}
}{fig1.1}{}
\*\*
\line{\vtop{\line{\hskip2.5truecm\vbox{\advance\hsize by -5.1 truecm
\0{{\css Fig. 1$\scriptstyle '$.}
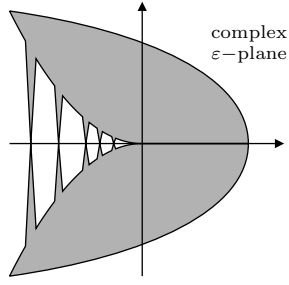
{\nota The domain $D_{0}$ of Figure 1 can be further extended? the
conjecture above asks whether the extended analyticity domain
could possibly be represented (close to the origin) as here: the
domain reaches the real axis at cusp points which are in $I_{\e_0}$
and correspond, in the complex $\e$-plane, to the elliptic tori which
are the analytic continuations of the hyperbolic tori. The analytic
continuation would be continuous thorough the real axis at the points of
$I_{\e_{0}}$. The cusps would be at least quadratic.\vfill}}
} \hfill} }}
\*
\endinsert

\*\*
\section(2,Formal solubility of the equations of motion)

\sub(2.1) {\it Formal expansion and recursive equations.}
We look for a formal expansion
$$ \eqalign{
\aaa(\pps;\e) & =
\sum_{k=1}^{\io} \e^{k} \aaa^{(k)}(\pps) =
\sum_{\nn\in\zzz^{r}} e^{i\nn\cdot\pps} \aaa_{\nn}(\e) =
\sum_{k=1}^{\io} \e^k \sum_{\nn\in \zzz^{r}}
e^{i\nn\cdot\pps} \, \aaa^{(k)}_{\nn} , \cr
\bbb(\pps;\e) & =
\sum_{k=1}^{\io} \e^{k} \bbb^{(k)}(\pps) =
\sum_{\nn\in\zzz^{r}} e^{i\nn\cdot\pps} \bbb_{\nn}(\e) =
\sum_{k=1}^{\io} \e^k \sum_{\nn\in \zzz^{r}}
e^{i\nn\cdot\pps} \, \bbb^{(k)}_{\nn} , \cr}
\Eq(2.1) $$
where we have not explicitly written the dependence on $\bb_{0}$.

Then to order $k$ the equations of motion \equ(1.6) become
$$ \eqalign{
\left( \oo\cdot\nn \right)^2 \aaa^{(k)}_{\nn} &
= \left[ \dpr_{\aa}f \right]^{(k-1)}_{\nn} , \cr
\left( \oo\cdot\nn \right)^2 \bbb^{(k)}_{\nn} &
= \left[ \dpr_{\bb}f \right]^{(k-1)}_{\nn} , \cr} 
\Eq(2.2) $$
where, given any function $F$ admitting a formal expansion
$$ F(\pps;\e) = \sum_{k=1}^{\io} \e^k\,\sum_{\nn\in\zzz^{r}}
e^{i\nn\cdot\pps} F^{(k)}_{\nn} ,
\Eq(2.3) $$
we denote by $[F]^{(k)}_\nn$ the coefficient with Taylor label $k$
and Fourier label $\nn$.

We can write
$$ \left[ \dpr_{\aa} f \right]^{(k-1)}_{\nn} =
\sum_{p \ge 0} \sum_{q \ge 0} {1\over p!} {1\over q!}
{\sum}^{*} \left( i\nn_0 \right)^{p+1}
\dpr_{\bb}^{q} f_{\nn_0}(\bb_{0})
\Big( \prod_{j=1}^{p} \aaa^{(k_{j})}_{\nn_{j}} \Big)
\Big( \prod_{j=p+1}^{p+q} \bbb^{(k_{j})}_{\nn_{j}} \Big) ,
\Eq(2.4) $$
where $0<k_{j}<k$ for all $j=1,\ldots,p+q$ and the $*$ denotes
that the sum has to be performed with the constraints
$$ 1 + \sum_{j=1}^{p+q} k_{j} = k , \qquad
\nn_{0} + \sum_{j=1}^{p+q} \nn_{j} = \nn ,
\Eq(2.5) $$
and, analogously,
$$ \left[ \dpr_{\bb} f \right]^{(k-1)}_{\nn} =
\sum_{p \ge 0} \sum_{q \ge 0} {1\over p!} {1\over q!}
{\sum}^{*} \left( i\nn_0 \right)^{p}
\dpr_{\bb}^{q+1} f_{\nn_0}(\bb_{0})
\Big( \prod_{j=1}^{p} \aaa^{(k_{j})}_{\nn_{j}} \Big)
\Big( \prod_{j=p+1}^{p+q} \bbb^{(k_{j})}_{\nn_{j}} \Big) ,
\Eq(2.6) $$
with the same meaning of the symbols.
\*

\sub(2.2) {\it Tree formalism.}
By iterating \equ(2.2), \equ(2.4) and \equ(2.6), one finds that one
can represent graphically $\aaa^{(k)}_{\nn}$ and $\bbb^{(k)}_{\nn}$ in
terms of {\it trees.}  The definition and usage of graphical tools
based on tree graphs in the context of KAM theory has been advocated
recently in the literature as an interpretation of the work \cita{E};
see for instance \cita{G1}, \cita{GG}, \cita{BGGM} and \cita{BaG}.

A tree $\th$ (see Figure 2 below) is defined as a partially ordered set
of points, connected by {\it lines}. The lines are
oriented toward the {\it root}, which is the leftmost point; the line
entering the root is called the {\it root line}. If a line $\ell$
connects two points $v_{1}$ and $v_{2}$ and is oriented from $v_2$ to
$v_1$ we say that $v_{2} \prec v_{1}$ and we shall write
$v_{1}'=v_{2}$ and $\ell=\ell_{v_{2}}$; we shall say also that $\ell$
exits from $v_{2}$ and enters $v_{1}$.

\midinsert
\*
\eqfig{200pt}{141.6pt}{
\ins{-29pt}{75pt}{\rm root}
\ins{14pt}{71pt}{$\nn\kern-2pt=\kern-2pt\nn_{\ell_{0}}$}
\ins{24pt}{91pt}{$\ell_{0}$}
\ins{50pt}{71pt}{$v_0$}
\ins{46pt}{91pt}{$\nn_{v_0}$}
\ins{127pt}{100.pt}{$v_1$}
\ins{121.pt}{125.pt}{$\nn_{v_1}$}
\ins{92.pt}{41.6pt}{$v_2$}
\ins{158.pt}{83.pt}{$v_3$}
\ins{191.7pt}{133.3pt}{$v_5$}
\ins{191.7pt}{100.pt}{$v_6$}
\ins{191.7pt}{71.pt}{$v_7$}
\ins{191.7pt}{-8.3pt}{$v_{11}$}
\ins{191.7pt}{16.6pt}{$v_{10}$}
\ins{166.6pt}{42.pt}{$v_4$}
\ins{191.7pt}{54.2pt}{$v_8$}
\ins{191.7pt}{37.5pt}{$v_9$}
}{fig2}{}
\*\*
\line{\vtop{\line{\hskip2.5truecm\vbox{\advance\hsize by -5.1 truecm
\0{{\css Fig. 2.}
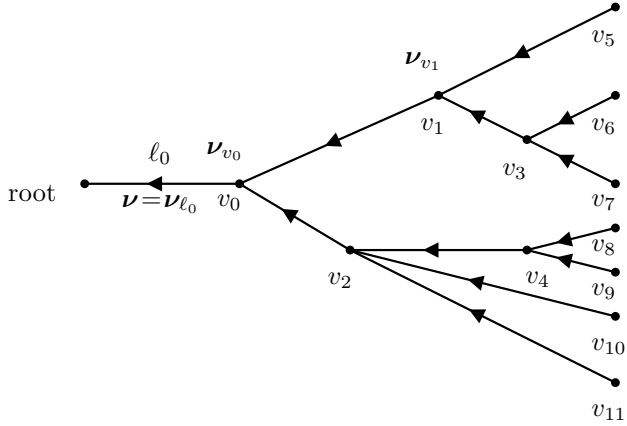
{\nota A tree $\st \th$ with $\st 12$ nodes;
one has $\st p_{v_0}=2,p_{v_1}=2,p_{v_2}=3,p_{v_3}=2,p_{v_4}=2$.
The length of the lines should be the same but it is drawn
of arbitrary size.\vfill}}
} \hfill} }}
\*
\endinsert

There will be two kinds of points: the {\it nodes} and the {\it leaves}.
The leaves can only be endpoints, \ie they have no lines
entering them, but an endpoint can be either a node or a leaf.  The
lines exiting from the leaves play a very different r\^ole with
respect to the lines exiting from the nodes, as we shall see below.
We shall denote by $v_{0}$ the last (\ie leftmost) node of the tree,
and by $\ell_{0}$ the root line; for future convenience we
shall write $v_{0}'=r$ but $r$ will not be considered a node.

We shall denote by $V(\th)$ the set of nodes,
by $L(\th)$ the set of leaves and by
$\L(\th)$ the set of lines.

Fixed any $\ell_{v} \in \th$, we shall say that the subset of $\th$
containing $\ell_{v}$ as well as all nodes $w\preceq v$ and all lines
connecting them is a {\it subtree} of $\th$ with root $v'$: of course
a subtree is a tree.

Given a tree, with each node $v$ we associate
a {\it mode label} $\nn_{v}\in\ZZZ^{r}$, and to each leaf $v$
a {\it leaf label} $\k_{v}\in \NNN$. The quantity
$$ k = |V(\th)| + \sum_{v\in L(\th)} \k_{v}
\Eq(2.7) $$
is called the {\it order} of the tree $\th$.

With any line $\ell$ exiting from a node $v$ we associate
two labels $\g_{\ell}$, $\g_{\ell}'$ assuming
the symbolic values $\a,\b$ and
a {\it momentum label} $\nn_{\ell}\in\ZZZ^{r}$,
which is defined as
$$ \nn_{\ell} \= \nn_{\ell_v} = \sum_{w\in V(\th) \atop
w \preceq v} \nn_{w} ,
\Eq(2.8) $$
while with any line $\ell$ exiting from a leaf $v$ we associate only the
labels $\g_{\ell}=\g_{\ell}'=\b$.

We can associate with each node also some labels depending on the
entering lines and on the exiting one:
the {\it branching labels} $p_{v}$ and $q_{v}$,
denoting how many lines $\ell$ having the label $\g_{\ell}=\a$
and, respectively, $\g_{\ell}=\b$ enter $v$,
and the label $\d_{v}$, defined as
$$ \d_{v} = \cases{
1 , & if $\g_{\ell_{v}} = \b $ , \cr
0 , & if $\g_{\ell_{v}} = \a $ . \cr}
\Eq(2.9) $$

Then with each node $v$ we associate a {\it node factor}
$$ F_{v} = {1\over p_{v}!} {1\over q_{v}!}
\Big( i\nn_{v} \Big)^{p_{v}+(1-\d_{v})}
\dpr_{\bb}^{q_{v}+\d_{v}} f_{\nn_v}(\bb_{0}) ,
\Eq(2.10) $$
which is a tensor of rank $p_{v}+q_{v}+1$,
while with each leaf $v$ we associate a {\it leaf factor}
(to be defined recursively, see below)
$$ L_{v} = \bbb^{(\k_{v})}_{\V0} ,
\Eq(2.11) $$
which is a tensor of rank 1 (\ie a vector);
to each line $\ell$ exiting from a node $v$ we associate
a {\it propagator}
$$ G_{\ell} \= \d_{\g_{\ell},\g_{\ell'}}
{ \openone \over (\oo\cdot\nn_{\ell})^2 } ,
\Eq(2.12) $$
which is a (diagonal) $d\times d$ matrix,
while no small divisor is associated with the lines
exiting from the leaves. For consistence we can define 
$$ G_{\ell} \= \d_{\g_{\ell},\g_{\ell'}}\,
\d_{\g_{\ell'},\b} \, \openone ,
\Eq(2.13) $$
for lines exiting from leaves, so that a propagator $G_{\ell}$
is in fact associated with each line.

\*

\sub(2.3) {\it Remark.}
Note that we can write \equ(2.12) in the form
$$ G_{\ell} = \Big( \matrix{
G_{\ell,\a\a} & G_{\ell,\a\b} \cr
G_{\ell,\b\a} & G_{\ell,\b\b} \cr} \Big)
\Eq(2.14) $$
where $G_{\ell,\a\a}$, $G_{\ell,\a\b}$, $G_{\ell,\b\a}$
and $G_{\ell,\b\b}$ are $r\times r$, $r\times s$, $s\times r$
and $s\times s$ matrices. By construction one has
$G_{\ell,\a\b}=G_{\ell,\b\a}^{T}=0$, and
$$ G_{\ell} = G(\oo\cdot\nn_{\ell}) , \qquad
G^{T}(-x) = G^{\dagger}(x) = G(x) ;
\Eq(2.15) $$
here and henceforth $T$ and $\dagger$ denote,
respectively, the transposed and the adjoint of a matrix.

\*

\sub(2.4) {\it Tree values and reduced tree values.}
Call $\Th_{k,\nn,\g}$ the set of all trees of order $k$ with
$\nn_{\ell_{0}}=\nn$ and $\g_{\ell_{0}}=\g$, if $\ell_{0}$ is the root
line. Set
$$ d_{\g} = \cases{ r , & for $\g=\a$ , \cr
s , & for $\g=\b$ ; \cr}
\Eq(2.16) $$
we can define an application $\Val\!:
\Th_{k,\nn,\g}\to \RRR^{d_{\g}}$, defined as
$$ \Val(\th) =
\Big( \prod_{v\in V(\th)} F_{v} \Big)
\Big( \prod_{v\in L(\th)} L_{v} \Big)
\Big( \prod_{\ell \in \L(\th)} G_{\ell} \Big) ,
\Eq(2.17) $$
which is called the {\it value} of the tree $\th$.


We can define also
$$ {\Val}'(\th) =
\Big( \prod_{v\in V(\th)} F_{v} \Big)
\Big( \prod_{v\in L(\th)} L_{v} \Big)
\Big( \prod_{\ell \in \L(\th) \setminus \ell_{0}} G_{\ell} \Big) ,
\Eq(2.18) $$
where, as usual, $\ell_{0}$ denotes the root line;
${\Val}'(\th)$ is called the {\it reduced value} of the tree $\th$.

The following cancellation is proved in Appendix \secc(A1).

\*

\sub(2.5) {\cs Lemma.} {\it Suppose that for all trees
$\th\in\Th_{k,\nn,\g}$ the set $\L(\th) \setminus \ell_{0}$
does not contain any lines $\ell$ with momentum $\nn_{\ell} = \V0$.
Then ${\Val}'(\th)$ is well defined and
$$ \sum_{\th\in\Th_{k,\V0,\a}} {\Val}'(\th) = \V0 .
\Eq(2.19) $$
}

\sub(2.6) {\it Existence of formal solutions.}
The following result states the existence of formal solutions
to \equ(1.6) which are conjugated to the unperturbed
motion \equ(1.5), provided the value $\bb_{0}$ is suitably fixed.

\*

\sub(2.7) {\cs Lemma.} {\it One can write, formally,
for all $\nn\in\ZZZ^{r}\setminus\{\V0\}$,
$$ \eqalign{
\aaa^{(k)}_{\nn} & = \sum_{\th\in \Th_{k,\nn,\a} }
\Val(\th) , \cr
\bbb^{(k)}_{\nn} & = \sum_{\th\in \Th_{k,\nn,\b} } \Val(\th) , \cr}
\Eq(2.20) $$
while $\aaa^{(k)}_{\V0}\=\V0$ and
$$ \bbb^{(k)}_{\V0} = -
\Big[ \dpr_{\bb}^{2} f_{\V0}(\bb_{0}) \Big]^{-1}
\sum_{\th\in \Th^{*}_{k+1,\V0,\b} } \Val'(\th) ,
\Eq(2.21) $$
where the quantities $\Val(\th)$ and $\Val'(\th)$ are
defined by \equ(2.17) and \equ(2.18), respectively,
and $*$ imposes the constraint that the tree whose reduced value
is given by $\dpr_{\bb}^{2}f_{\V0}(\bb_{0})\,\bbb^{(k)}_{\V0}$
has to be discarded from the set $\Th_{k+1,\V0,\b}$.
If one has
$$ \eqalign{
& \dpr_{\bb}f_{\V0}(\bb_{0}) = \V0 , \cr
& \det \dpr_{\bb}^{2} f_{\V0}(\bb_{0}) \neq 0 , \cr}
\Eq(2.22) $$
then there exists a unique way to fix $\bbb^{(k)}_{\V0}$
for all $k\in\NNN$ such that $\aaa^{(k)}_{\nn}$
and $\bbb^{(k)}_{\nn}$ are finite for all
$\nn\in\ZZZ^{p}\setminus\{\V0\}$ to all perturbative orders $k$.}

\*

\sub(2.8) {\it About the proof.}
The proof of \equ(2.20) is by induction.
In order to show that it is possible to fix uniquely $\bbb^{(k)}_{\V0}$
so that the existence of a formal solution follows,
the key is to realize that no division by zero occurs
in the recursive solution of \equ(2.2): the
coefficients $\bbb^{(k)}_{\V 0}$ are determined precisely by imposing
the validity of this property for the lines $\ell$
with $\g_{\ell}=\g_{\ell}'=\b$. In fact the condition to avoid dividing
by zero takes, to all orders $k$ in $\e$, the form $\dpr_{\bb}^{2}
f_{\V0}(\bb_0)\,\bbb^{(k)}_{\V 0}=$ some vector
determined recursively, so that $\bbb^{(k)}_{\V0}$
is defined by exploiting the assumption \equ(2.21). A
further key point is to realize that the lines $\ell$ with
$\g_{\ell}=\g_{\ell}'=\a$ and carrying $\nn_{\ell}=\V0$ never appear,
and the previous lemma is enough to imply this. Details of the proof
are given in Appendix \secc(A2). \qed

\*

\sub(2.9) {\it Remark}. By \equ(2.2) and by the lemma \secc(2.7)
one has
$$ \eqalign{
\Big[ \dpr_\aa f \Big]^{(k)}_{\nn} & = \sum_{\th\in \Th_{k,\nn,\a} }
{\Val}'(\th) , \cr
\Big[ \dpr_\bb f \Big]^{(k)}_{\nn} & = \sum_{\th\in \Th_{k,\nn,\b} }
{\Val}'(\th) , \cr}
\Eq(2.23) $$
as one realizes by comparing \equ(2.17) with \equ(2.18).

\*

\sub(2.10) {\it Remark.} As it will follow from the analysis
performed in the next Sections, the tools described above
are sufficient to prove the convergence (hence the analyticity)
of the perturbative expansions \equ(2.1), for $\e$ small enough:
in fact we shall see that the main techniacl difficulties shall arise
from the problem of bounding the propagators, while,
in the case of periodic solutions (\ie $r=1$) we can simply
bound $G_{\ell}$ by the inverse of the rotation vector $\o$
(which is a number in such a case).

\*\*
\section(3,Self-energy graphs)

\sub(3.1) {\it Trimmed trees.}
With respect to the papers \cita{GG}, \cita{BGGM} and \cita{BaG}, the
trees here carry also ``leaves'':
each leaf can be decomposed in terms of trees, because
$\bbb^{(k)}_{\V0}$ is given by
$$ \bbb^{(k)}_{\V0} = -
\Big[ \dpr_{\bb}^{2} f_{\V0}(\bb_{0}) \Big]^{-1} \GG_{\V0}^{(k+1)} ,
\Eq(3.1) $$
with $\GG^{(k+1)}_{\V0}$ expressed as sum of reduced values
of trees of order $k+1$ (see Appendix \secc(A2)); more precisely
$$ \GG^{(k+1)}_{\V0} =
\sum_{\th\in\Th_{k+1,\V0,\b}} {\Val}'(\th) -
\dpr_{\bb}^{2}f_{\V0}(\bb_{0}) \bbb^{(k)}_{\V0} \=
\sum_{\th\in\Th^{*}_{k+1,\V0,\b}} {\Val}'(\th) ,
\Eq(3.2) $$
where $*$ has been define after \equ(2.21):
it recalls that the tree whose reduced value is given
by $\dpr_{\bb}^{2} f_{\V0}(\bb_{0})\,\bbb^{(k)}_{\V0}$
has to be discarded from the set $\Th_{k+1,\V0,\b}$.

Of course each leaf can contain other leaves and so on.
If each time a leaf is encountered, it can be decomposed into trees,
at the end we have that the value of a tree $\th$
can be expressed as product of factors
which are values of trees without leaves, that we can call,
as in \cita{Ge}, {\it trimmed trees}.
The sum of the orders of all the so obtained
trimmed trees is equal to $k$, if the tree $\th$ belonged to
$\Th_{k,\nn,\g}$; moreover for all trimmed trees the
order equals exactly the number of nodes, as it follows
from \equ(2.7) by using that a trimmed tree has no leaves.

\*

\sub(3.2) {\it Multi-scale decomposition and clusters.}
Given a vector $\oo\in D_{\t}(C_{0})$, define
$\oo_{0}=2^{\t}C_{0}^{-1}\oo$.
Then there exists a sequence $\{\g_{n}\}_{n\in\zzz_{+}}$,
with $\g_{n}\in[2^{n-1},2^{n}]$, such that
$$ \left| \left| \oo_{0}\cdot\nn \right| - \g_{p} \right|
\ge 2^{n+1} \qquad \hbox{ if } 0 <
\left| \nn \right| \le 2^{-(n+3)/\t} ,
\Eq(3.3) $$
for all $n\le0$ and for all $p\ge n$, and $|\oo_{0}\cdot\nn|\neq\g_{n}$
for all $\nn\in\ZZZ$ and for all $n\le 0$;
the existence of such a sequence (depending on $\oo$)
is proved by Proposition in Section 3 of \cita{GG}.

Given a line $\ell$ with momentum $\nn_{\ell}$
we say that $\ell$ has scale label $n_{\ell}=1$ if
$$ \left| \oo_{0}\cdot\nn_{\ell} \right| \ge \g_{0} ,
\Eq(3.4) $$
and scale label $n_{\ell}=n \in \ZZZ\setminus\ZZZ_{+}$ if
$$ \g_{n-1} \le \left| \oo_{0}\cdot\nn_{\ell} \right| < \g_{n} .
\Eq(3.5) $$

Once the scale labels have been assigned to the lines one has a
natural decomposition of the tree into clusters.  A {\it cluster} $T$
on scale $n$ is a maximal set of nodes and lines connecting them such
that all the lines have scales $n'\ge n$ and there is at least one
line with scale $n$; if a cluster $T'$ is contained inside a cluster $T$
we shall say that $T'$ is a subcluster of $T$.
The $m_{T}\ge 0$ lines entering the cluster $T$ and the possible
exiting line (unique if existing at all) are called the
{\it external lines} of the cluster $T$; given a cluster $T$
on scale $n$, we shall denote by $n_{T}=n$ the scale of the cluster.
We call $\TT(\th)$ the set of all clusters in a tree $\th$.  

Given a cluster $T\in \TT(\th)$ call $V(T)$, $L(T)$ and $\L(T)$
the set of nodes, the set of leaves and the set of lines of $T$,
respectively. Let us define also
$$ \nn_{T} = \sum_{v\in V(T)} \nn_{v} ,
\Eq(3.6) $$
and denote by $\TT_{\V0}(\th)$ the set of all clusters
$T$ with $\nn_{T}=\V0$.
Given a cluster $T$ call $T_{0}$ the subset of $T$ obtained
from $T$ by eliminating all the nodes and lines
of the subclusters $T' \subset T$ such that $\nn_{T'}=\V0$,
and denote by $V(T_{0})$ and $\L(T_{0})$ the set of nodes
and lines, respectively, in $T_{0}$.

\*

\sub(3.3) {\it Self-energy graphs.}
We call {\it self-energy graphs} of a tree $\th$ the clusters $T\in
\TT(\th)$ such that
\\
(1) $T$ has only one entering line $\ell_{T}^{2}$
and one exiting line $\ell_{T}^{1}$,
\\
(2) $T\in \TT_{\V0}(\th)$, \ie
$$ \nn_{T} \= \sum_{v\in V(T)} \nn_{v} = \V0 ,
\Eq(3.7) $$
(3) one has
$$ \sum_{v\in V(T_{0})} |\nn_{v}| \le 2^{-(n+3)/\t} ,
\Eq(3.8) $$
where $n_{\ell_{T}^{2}} = n$ is the scale of the line $\ell_{T}^{2}$.

We say that the line $\ell_{T}^{1}$ exiting a self-energy graph $T$ is
a {\it self-energy line}; we call {\it normal line} any line of the
tree which is not a self-energy line.

Given a self-energy graph $T\in \TT(\th)$ we say that a
self-energy graph $T'\in \TT(\th)$ contained in $T$ is
{\it maximal} if there are no other self-energy graphs
internal to $T$ and containing $T'$.
We say that a self-energy graph $T$ has {\it height} $D_{T}=0$
if it does not contain any other self-energy graphs,
and that it has height $D_{T}=D\in \ZZZ_{+}$, recursively,
if it contains maximal self-energy graphs with height $D-1$.

\midinsert
\*
\eqfig{160pt}{140pt}{
\ins{55pt}{60pt}{$v_{1}$}
\ins{110pt}{71pt}{$v_{2}$}
\ins{69pt}{93pt}{$v_{3}$}
\ins{134pt}{78pt}{$v_{5}$}
\ins{115pt}{117pt}{$v_{6}$}
\ins{120pt}{86pt}{$v_{4}$}
\ins{10pt}{80pt}{$\ell^1_T$}
\ins{85pt}{33pt}{$T$}
\ins{130pt}{110pt}{$T'$}
\ins{100pt}{93pt}{$T''$}
\ins{190pt}{58pt}{$\ell^2_T$}
}{fig3}{}
\*\*
\line{\vtop{\line{\hskip2.5truecm\vbox{\advance\hsize by -5.1 truecm
\0{{\css Fig. 3.}
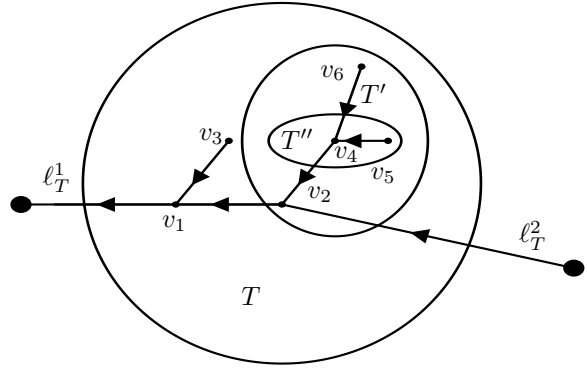
{\nota Three self-energy graphs $T,T',T''$ of height $2$,
$1$, $0$, respectively.
The black balls represent the remaining parts of the trees;
the labels are not explicitly shown.
The self-energy graph $T'$ is maximal in $T$.
The ellipses are visual aids to identify the clusters.
The scales (not marked) of the lines increase as one crosses inward
the ellipses boundaries: recall, however, that the scale labels are $\le0$.
The cluster $T_{0}$, defined after \equ(3.6),
consists in this case of the nodes $v_{1}$ and $v_{3}$,
and of the lines $\ell_{v_{3}}$ and $\ell_{v_{2}}$.
The mode labels of the pairs of nodes $(v_{1},v_{3})$,
$(v_{2},v_{6})$ and $(v_{4},v_{5})$ add up to $\V0$.
An example of a simpler self-energy graph
could be obtained by deleting the line $\ell_{v_{3}}$
and assuming that the mode label of $v_{1}$ vanishes.\vfill}}
} \hfill} }}
\*
\endinsert

Given a line $\ell\in \L(T_{0})$ with momentum $\nn_{\ell}$,
its {\it reduced momentum} $\nn_{\ell}^{0}$ is defined as
$$ \nn_{\ell}^{0} = \sum_{w \in V(T_{0}) \atop w\preceq v} \nn_{w} ,
\qquad \ell \= \ell_{v} ,
\Eq(3.9) $$
and it can be given a scale $n_{\ell}^{0}$ such that
$$ \g_{n_{\ell}^{0}-1} \le
\left| \oo_{0}\cdot\nn_{\ell}^{0} \right| < \g_{n_{\ell}^{0}} ;
\Eq(3.10) $$
we call $n_{\ell}^{0}$ the {\it reduced scale} of the line $\ell$.

\*

\sub(3.4) {\it Remarks.}
(1) Given a self-energy  graph $T$, for all lines $\ell\in \L(T)$,
one can write, by setting $\ell=\ell_{v}$,
$$ \nn_{\ell}=\nn_{\ell}^{0} + \s_{\ell}\nn ,
\Eq(3.11) $$
where $\nn\=\nn_{\ell_{T}^{2}}$ is the momentum flowing
through the line $\ell_{T}^{2}$ entering $T$,
while $\s_{\ell}$ is defined as follows: writing $\ell\=\ell_{v}$
then $\s_{\ell}=1$ if $\ell_{T}^{2}$ enters a node $w\preceq v$
and $\s_{\ell}=0$ otherwise.
\\
(2) Note that the entering line $\ell_{T}^{2}$ must have,
by the condition \equ(3.7), the same momentum
as the exiting line $\ell_{T}^{1}$, hence,
by construction, the same scale $n_{\ell_{T}^{2}}=n_{\ell_{T}^{1}}$.
\\ 
(3) The notion of self-energy lines has been introduced by Eliasson
who named them ``resonances'', \cita{E}. We change the name here not only
to avoid confusion with the notion of mechanical resonance (which is
related to a rational relation between frequencies of a quasi periodic
motion) but also because the ``tree expansions'' that we use here (also
basically due to Eliasson) can be interpreted, \cita{GGM}, as Feynman
graphs of a suitable field theory. As such they correspond to classes
of self-energy graphs: we use here the correspondence to perform
resummation operations typical of renormalization theory.

\*

\sub(3.5) {\it Value of a self-energy graph.}
Given a self-energy  graph $T$ and denoting by $|V(T)|$
the number of nodes in $T$, define the {\it self-energy value} as
$$ \VV_{T}(\oo\cdot\nn) =
\e^{|V(T)|} \Big( \prod_{v\in V(T)} F_{v} \Big)
\Big( \prod_{v\in L(T)} L_{v} \Big)
\Big( \prod_{\ell\in \L(T)} G_{\ell} \Big) ,
\qquad |V(T)| \ge 1 ,
\Eq(3.12) $$
seen as a function of $\oo\cdot\nn$, if $\nn\=\nn_{\ell_{T}^{2}}=
\nn_{\ell_{T}^{1}}$ is the momentum flowing through
the external lines of the self-energy  graph $T$.
Recall that we are considering trimmed trees, so that
no leaves can appear; see \sec(3.1).

We can have four types of self-energy graphs depending on the types $\a$ or
$\b$ of the labels $\g_{\ell^{1}_{T}}'$ and $\g_{\ell^{2}_{T}}$:
$$ \matrix{
& \qquad \g_{\ell_{T}^{1}}' & \qquad \g_{\ell_{T}^{2}} \cr
& & \cr
1. & \qquad \a & \qquad \a \cr
2. & \qquad \a & \qquad \b \cr
3. & \qquad \b & \qquad \a \cr
4. & \qquad \b & \qquad \b \cr}
\Eq(3.13) $$

Given a tree $\th$, define
$$
N_{n}(\th)
= \left\{ \ell \in \L(\th) \; : \; n_{\ell} = n \right\} ,
\Eq(3.14) $$
and
$$ M(\th) = \sum_{v \in V(\th)} \left| \nn_{v} \right| .
\Eq(3.15) $$
Call $N_{n}^{*}(\th) $ the number of normal lines on scale
$n$ and call $R_{n}(\th)$ the number of self-energy lines
on scale $n$. Of course
$$ N_{n}(\th) = N_{n}^{*}(\th) + R_{n}(\th) .
\Eq(3.16) $$

Then the following result holds;
it is a version of the key estimate of Siegel's theory in the
interpretation of Bryuno \cita{B} and P\"oschel \cita{P}.
This is proved as in \cita{G1} or \cita{BaG}, for insatnce;
howeve, for completeness, a proof is also given in Appendix \secc(A3).

\*

\sub(3.6) {\cs Lemma.}
{\it For any tree $\th\in\Th_{\k,\nn,\g}$ one has
$$ N_{n}^{*}(\th)  
\le c\, M(\th) \, 2^{n/\t} ,
\Eq(3.17) $$
for some constant $c$.}

\*

\sub(3.7) {\it Localization operators.}
For any self-energy  graph $T$ we define
%
%
%
$$
\LL \VV_{T}(\oo\cdot\nn)
\= \VV_{T}(0) + \left( \oo\cdot\nn \right) \dpr\VV_{T}(0) , 
\Eq(3.19) $$
were $\dpr\VV_{T}$ denotes the first derivative of $\VV_{T}$
with respect to its argument;
the quantity $\VV_{T}(0)$ is obtained from
$\VV_{T}(\oo\cdot\nn)$ by replacing $\nn_{\ell}$ with
$\nn_{\ell}^{0}$ in the argument of each propagator
$G_{\ell}$, while $\dpr\VV_{T}(0)$ is obtained from
$\VV_{T}(\oo\cdot\nn)$ by differentiating it with
respect to $x=\oo\cdot\nn$, and thence
replacing $\nn_{\ell}$ with $\nn_{\ell}^{0}$
in the argument of each propagator $G_{\ell}$.

We shall call $\LL$ the {\it localization operator} and 
$\LL \VV_{T}(\oo\cdot\nn)$
the {\it localized part}
of the self-energy value.

%
%

\*

\sub(3.8) {\it Families of self-energy graphs.}
Given a tree $\th$ containing a self-energy graph $T$, we can consider
all trees obtained by changing the location of the nodes in $T_{0}$
(note that $T_0$ is defined after \equ(3.6)) which the external
lines of $T$ are attached to: we denote by $\FF_{T_{0}}(\th)$ the set
of trees so obtained, and call it {\it the self-energy family}
associated with the self-energy graph $T$.  And we shall refer to the
operation of detaching and reattaching the external lines, by saying
that we are {\it shifting} such lines.

Of course shifting the external lines of a self-energy  graph produces a
change of the propagators of the trees.
In particular since all arrows have to point toward the root,
some lines can revert their arrows.

Moreover the momentum can change,
as a reversal of the arrow implies a change of the
partial ordering of the nodes inside the self-energy graph 
and a shifting of the entering line can add or subtract
the contribution of the momentum flowing through it.
More precisely, if the external lines of a self-energy graph  $T$ are
detached then reattached to some other nodes in $V(T)$,
the momentum flowing through the line $\ell\in \L(T)$ can be
changed into $\pm\nn_{\ell}^{0}+\s \nn$, with $\s\in\{0,1\}$:
if we call $V_{1}$ and $V_{2}$ the two disjoint sets
into which $\ell$ divides $T$ (see Figure 2), such that the arrow
superposed on $\ell$ is directed from $V_{2}$ to $V_{1}$
(before detaching the external lines), then
the sign is $+$ if the exiting line is reattached to
a node inside $V_{1}$ and it is $-$ otherwise,
while $\s=1$ if the entering line is reattached to
a node inside $V_{2}$ when the sign is $+$ and
to a node inside $V_{1}$ when the sign is $-$,
and $\s=0$ otherwise.

\*
\eqfig{200pt}{150pt}{
\ins{85pt}{115pt}{$V_{1}$}
\ins{140pt}{40pt}{$V_{2}$}
\ins{46pt}{105pt}{$v_{1}$}
\ins{75pt}{133pt}{$v_{2}$}
\ins{106pt}{45pt}{$v_{3}$}
\ins{117pt}{74pt}{$v_{4}$}
\ins{142pt}{58pt}{$v_{5}$}
\ins{115pt}{32pt}{$v_{6}$}
\ins{10pt}{125pt}{$\ell_{T}^{1}$}
\ins{185pt}{65pt}{$\ell_{T}^{2}$}
\ins{80pt}{65pt}{$\ell$}
\ins{140pt}{120pt}{$T$}
}{fig4}{}
\*\*
\line{\vtop{\line{\hskip2.5truecm\vbox{\advance\hsize by -5.1 truecm
\0{{\css Fig. 4.}
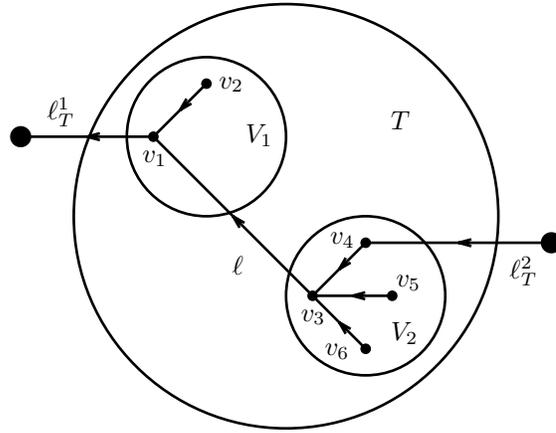
{\nota The sets $\st V_{1}$ and $\st V_{2}$
in a self-energy graph $\st T$; note that, even if they
are drawn like circles, the sets $\st V_{1}$ and
$\st V_{2}$ are not clusters.
One has $\st \nn_{\ell}^{0}=
\nn_{v_{3}}+\nn_{v_{4}}+\nn_{v_{5}}+\nn_{v_{6}}$ and
$\st \nn=\nn_{\ell_{T}^{2}}$; of course
$\st \nn_{\ell_{T}^{1}}=\nn_{\ell_{T}^{2}}$
and $\st \nn_{\ell}^{0}=-(\nn_{v_{1}}+\nn_{v_{2}})$
by definition of rself-energy graph.
The black balls represent the remaining parts of the trees.
The labels are not explicitly shown.\vfill}}
} \hfill} }}
\*

Referring to \equ(3.13) for the notion of type of self-energy graph
one shows the existence of the following cancellations
(the proof is in Appendix \secc(A4)).
\*

\sub(3.9) {\cs Lemma.}
{\it Given a tree $\th$, for any self-energy  graph
$T\in \TT(\th)$ one has
$$ \sum_{\th'\in \FF_{T_{0}}(\th)} \LL\VV_{T}(\oo\cdot\nn) =
\cases{
0 , & if $T$ is of type 1 , \cr
(\oo\cdot\nn) B_{\FF_{T_{0}}(\th)}' , & if $T$ is of type 2 , \cr
(\oo\cdot\nn) B_{\FF_{T_{0}}(\th)}'' , & if $T$ is of type 3 , \cr
A_{\FF_{T_{0}}(\th)} , & if $T$ is of type 4 , \cr}
\Eq(3.24) $$
where $\nn=\nn_{\ell_{T}^{2}}$, the sum is over the
self-energy family associated with $T$, and
$A_{\FF_{T_{0}}(\th)}$, $B_{\FF_{T_{0}}(\th)}'$
and $B_{\FF_{T_{0}}(\th)}''$ are matrices
$s\times s$, $r\times s$ and $s\times r$,
respectively, depending only on the self-energy  graph $T$;
in particular they are independent of the quantity $\oo\cdot\nn$.}

\*

\sub(3.10) {\it First step toward the resummation of self-energy graphs.}
Let $\th$ be a tree $\th\in\Th_{\nn,k,\g}$ with a self-energy graph
$T$. Define $\th_{0}=\th\setminus T$ as the set of nodes and lines of
$\th$ outside $T$ (of course $\th_{0}$ is not a tree).  Consider
simoultaneously all trees such that the structure $\th_{0}$ outside of
the self-energy graph is the same, while the self-energy graph itself
can be arbitrary, \ie $T$ can be replaced by any other self-energy
graph $T'$ with $k_{T'} \ge 1$.  This allows us to define as a formal
power series the matrix
$$ M(\oo\cdot\nn;\e) =
\sum_{\th=\th_{0}\cup T'} \VV_{T'}(\oo\cdot\nn)
\Eq(3.25) $$
where the sum is over all trees $\th$ such that
$\th\setminus T$ is fixed to be $\th_{0}$
and the mode labels of the nodes $v\in V(T)$
have to satisfy the conditions $(1)\div(3)$ in \sec(3.3)
defining the self-energy graphs.

The following property holds (the proof is in Appendix \secc(A5)) as
an algebraic identity between formal power series.

\*

\sub(3.11) {\cs Lemma.} {\it
The following two properties hold:
(1) $\left( M(x;\e) \right)^{T} = M(-x;\e)$, and
(2) $\left( M(x;\e) \right)^{\dagger}$ $=$ $M(x;\e)$;
the latter means that the matrix $M(x;\e)$ is self-adjoint.}

\*

\sub(3.12) {\it Remarks.}
(1) The function $M(\oo\cdot\nn;\e)$ depends on $\e$ but, by
construction, it is independent of $\th_{0}$: hence we can rewrite
\equ(3.25) as
$$ M(\oo\cdot\nn;\e) = \sum_{T'} \VV_{T'}(\oo\cdot\nn) ,
\Eq(3.26) $$
where the sum is over all self-energy graphs of order $k\ge1$
with external lines with momentum $\nn$.
\\
(2) In \equ(3.25) or \equ(3.26),
if $\g_{n-1}\le |\oo_{0}\cdot\nn| < \g_{n}$,
the sum has to be restricted to the
self-energy graphs $T'$ on scale $n_{T'} \ge n+3$.
Writing, for any line $\ell\in T_{0}'$,
$\nn_{\ell}$ as in \equ(3.12) one has
$$ \left| \oo_{0}\cdot\nn_{\ell}^{0} \right| >
2^{\t}C_{0}^{-1} C_{0} \left| \nn_{\ell}^{0} \right|^{-\t} \ge
2^{\t} \Big( \sum_{v\in V(T_{0})} |\nn_{v}| \Big)^{-\t}
\ge 2^{\t}2^{n+3} ,
\Eq(3.27) $$
while $\left|\oo_{0}\cdot\nn \right|<2^{n}$, so that,
by using \equ(3.11), one obtains
$$ \left| \oo_{0}\cdot\nn_{\ell} \right|
> 2^{\t}2^{n+3} - 2^{n} > 2^{n+2} ,
\Eq(3.28) $$
which implies $n_{\ell}\ge n+3$.
\\
(3) The matrix $M(\oo\cdot\nn;\e)$ can be written as
$$ M(\oo\cdot\nn;\e) = \Big( \matrix{
M_{\a\a}(\oo\cdot\nn;\e) & M_{\a\b}(\oo\cdot\nn;\e) \cr
M_{\b\a}(\oo\cdot\nn;\e) & M_{\b\b}(\oo\cdot\nn;\e) \cr} \Big)
\Eq(3.29) $$
where $M_{\a\a}(\oo\cdot\nn;\e)$, $M_{\a\b}(\oo\cdot\nn;\e)$,
$M_{\b\a}(\oo\cdot\nn;\e)$ and $M_{\b\b}(\oo\cdot\nn;\e)$
are $r\times r$, $r\times s$, $s\times r$ and $s\times s$ matrices.
It is easy to realize that (up to convergence problems to be
discussed in Section \secc(5))
$$ \eqalign{
M_{\a\a}(\oo\cdot\nn;\e) & = O(\e^{2}(\oo\cdot\nn)^{2})) , \cr
M_{\a\b}(\oo\cdot\nn;\e) & = O(\e^{2}(\oo\cdot\nn)) , \cr
M_{\b\b}(\oo\cdot\nn;\e) & = O(\e)+ O(\e^{2}(\oo\cdot\nn)^{2}) . \cr}
\Eq(3.30) $$
The proportionality of $M_{\a\a}(\oo\cdot\nn;\e)$ to
$(\oo\cdot\nn)^2$ and of $M_{\a\b}(\oo\cdot\nn;\e)$ to $\oo\cdot\nn$
is a consequence of the lemma \secc(3.9).
First orders computations already give, for instance,
$$ \eqalign{
M_{\b\b}(\oo\cdot\nn;\e) & =
\e \dpr_{\bb}^{2}f_{\V0}(\bb_{0}) + O(\e^{2}) +
O(\e^{2}(\oo\cdot\nn)^{2}) , \cr
M_{\a\b} (\oo\cdot\nn;\e) & =
-2 \e^{2}i \left( \oo\cdot\nn \right) \sum_{\nn_{1}+\nn_{2} = \V0
\atop |\nn_{1}|+|\nn_{2}| < 2^{-(n+3)/\t} }
{1 \over (\oo\cdot\nn_{2})^{3} } \cr
& \qquad \Big[
\nn_{1}\dpr_{\bb}f_{\nn_{1}}(\bb_{0}) \dpr^{2}_{\bb}f_{\nn_{2}}(\bb_{0})
-\nn_{1}^{2} f_{\nn_{1}}(\bb_{0}) \nn_{2}\dpr_{\bb}f_{\nn_{2}}(\bb_{0})
\Big] + O(\e^{3}(\oo\cdot\nn)) , \cr}
\Eq(3.31) $$
where $\g_{n-1}\le |\oo_{0}\cdot\nn|<\g_{n}$.
Therefore $M_{\b\b}(\oo\cdot\nn;\e) \neq 0$
by hypothesis (see \equ(1.8)), and
$M_{\a\b}(\oo\cdot\nn;\e)$ is generically nonvanishing.
\\
(4) The lemma \secc(3.11) implies, that, by defining
the matrices
$B_{\FF_{T_{0}}(\th)}'$ and $B_{\FF_{T_{0}}(\th)}''$
as in the lemma \secc(3.9), one has
$$ B_{\FF_{T_{0}}(\th)}' = - \left( B_{\FF_{T_{0}}(\th)}'' \right)^{T} .
\Eq(3.32) $$

\*

\sub(3.13) {\it Changing scales.}
When shifting the lines external to the self-energy graphs,
the momenta of the internal lines can change.
As a consequence in principle also the scale labels could change;
however this does not happen, as the following result shows;
for the proof see Appendix \secc(A6).

\*

\sub(3.14) {\cs Lemma.}
{\it For all lines $\ell\in\L(\th)$ one has
$n_{\ell}=n_{\ell}^{0}$; in particular this implies that,
when shifting the lines external to the self-energy graphs
of a tree $\th$, the scale labels $n_{\ell}$ of all
line $\ell\in\L(\th)$ do not change.}

\*\*
\section(4,Resummations of self-energy graphs: renormalized propagators)

\sub(4.1) {\it Renormalized trees.}
So far we considered formal power expansions in $\e$.
By introducing the function $\hh=(\hh_{\a},\hh_{\b})=(\aaa,\bbb)$,
we can write the function $\hh(\pps,\bb_{0};\e)\=\hh(\pps;\e)$ as
$$ \hh(\pps;\e) = \sum_{\nn\in\zzz^{r}}
e^{i\nn\cdot\pps} \hh_{\nn}(\e) ,
\Eq(4.1) $$
because we are looking for a solution periodic in $\pps\in\TTT^{r}$.

In terms of the formal power expansion envisaged in Section \secc(3),
we can define a solution ``approximated to order $k$'' as
$$ \hh^{(\le k)}_{\nn}(\e) =
\sum_{k'=1}^{k} \e^{k'} \hh^{(k')}_{\nn} , \qquad
\hh^{(k')}_{\g\nn} = \sum_{\th\in\Th_{k',\nn,\g}} \Val(\th) ,
\Eq(4.2) $$
where $\Th_{k',\nn,\g}$ is defined in \sec(2.4),
and $\Val(\th)$ is given by \equ(2.17).

However we can define a different sequence of {\it approximating
functions} $\ohh^{[k]}(\pps;\e)$, formally converging to the formal
solution (as we shall see in the proposition \secc(5.13) below), by
defining it iteratively as follows.

Denote by $\Th^{\RR}_{k,\nn,\g}$ the set of all trees
of order $k$ without self-energy graphs and
with labels $\nn_{\ell_{0}}=\nn$ and $\g_{\ell_{0}}=\g$
associated with the root line; we shall call $\Th^{\RR}_{k,\nn,\g}$
the set of {\it renormalized trees} of order $k$
(and with labels $\nn$ and $\g$ associated with the root line).
Given a tree $\th \in \Th^{\RR}_{k,\nn,\g}$ and
a cluster $T\in \TT(\th)$, by extension we shall say that $T$
is a {\it renormalized cluster}.

We can also consider a self-energy graph which does not contain
any other self-energy graph: we shall say that
such a self-energy graph is a {\it renormalized self-energy graph};
of course no one of such clusters can appear in any tree
in $\Th^{\RR}_{k,\nn,\g}$.

For a renormalized tree $\th$ of arbitrary order $k'$, define
$$ \oV^{[k]}(\th) =
\Big( \prod_{v\in V(\th)} F_{v} \Big)
\Big( \prod_{v\in L(\th)} L_{v} \Big)
\Big( \prod_{\ell \in \L(\th)} \ol{G}_{\ell}^{[k-1]} \Big) ,
\Eq(4.3) $$
with the {\it dressed propagators} given by
$$ \cases{
\ol{G}_{\ell}^{[0]} = \left( \oo\cdot\nn_{\ell} \right)^{-2}
\openone \d_{\g_{\ell},\g_{\ell'}} , & \cr
\ol{G}_{\ell}^{[k]} = \left[
\left( \oo\cdot\nn_{\ell} \right)^{2} \openone -
M^{[k]}(\oo\cdot\nn_{\ell};\e) \right]^{-1} , &
for $k\ge 1 $ , \cr}
\Eq(4.4) $$
where the sequence $\{ M^{[k]}(\oo\cdot\nn;\e) \}_{k\in\nnn}$
is iteratively defined as sum of the values of all
renormalized self-energy graphs which can be
obtained by using the propagators $\ol{G}_{\ell}^{[k-1]}$, \ie as
$$ \eqalign{
& M^{[k]}(\oo\cdot\nn;\e) =
\sum_{{\rm renormalized}\;T} \VV^{[k]}_{T}(\oo\cdot\nn) , \cr
& \VV^{[k]}_{T}(\oo\cdot\nn) = 
\e^{|V(T)|} \, \Big( \prod_{v\in V(T)} F_{v} \Big)
\Big( \prod_{v\in L(T)} L_{v} \Big)
\Big( \prod_{\ell\in \L(T)} \ol{G}^{[k-1]}_{\ell} \Big) . \cr}
\Eq(4.5) $$
where $|V(T)|$ is the number of nodes in $T$;
we can also define $M^{[0]}(\oo\cdot\nn;\e)\=0$.
The leaf factors $L_{v}$ in \equ(4.3) are recursively defined as
$$ L_{v} = \bbb_{0}^{[k,\k_{v}]} =
- \left[ \dpr_{\bb}^{2}f_{\V0}(\bb_{0}) \right]^{-1}
\sum_{\th\in \Th^{\RR*}_{\k_{v},\V0,\b}} {\oV^{[k]}}'(\th) ,
\Eq(4.6) $$
where
$$ {\oV^{[k]}}'(\th) =
\Big( \prod_{v\in V(\th)} F_{v} \Big)
\Big( \prod_{v\in L(\th)} L_{v} \Big)
\Big( \prod_{\ell \in \L(\th)\setminus\ell_{0}}
\ol{G}_{\ell}^{[k-1]} \Big) ,
\Eq(4.7) $$
and $*$ has the same meaning as after \equ(3.2).

To avoid confusing the value of a renormalized tree with the
tree value introduced in \equ(2.13), we shall call \equ(4.3) the
{\it renormalized value} of the (renormalized) tree.

Then we shall write
$$ \eqalign{
& \ohh^{[k]}(\pps;\e) = \sum_{\nn\in\zzz^{r}}
e^{i\nn\cdot\pps} \ohh^{[k]}_{\nn}(\e) , \cr
& \ohh^{[k]}_{\nn}(\e) =
\sum_{k'=1}^{\io} \e^{k'} \ohh^{[k,k']}_{\nn}(\e) , \qquad
\ohh^{[k,k']}_{\g\nn}(\e) =
\sum_{\th\in\Th^{\RR}_{k',\nn,\g}} \oV^{[k]}(\th) , \cr}
\Eq(4.8) $$
where the last formula holds for $\nn\neq\V0$, because
for $\nn=\V0$, one has \equ(4.6) for $\g_{\ell}=\b$,
while $\ohh^{[k,k']}_{\g\V0}\=0$.

\*

\sub(4.2) {\it Remark.}
Note that if we expand the quantity $M^{[k]}(\oo\cdot\nn;\e)$
in powers of $\e$, by expanding the propagators
$\ol{G}^{[k-1]}_{\ell}$, we reconstruct the sum
of the values of all self-energy graphs containing only
self-energy graphs with height $D \le k$. Therefore
if we expand $M^{[k+1]}(\oo\cdot\nn;\e)$ in powers of $\e$
we obtain the same terms as if expanding
$M^{[k]}(\oo\cdot\nn;\e)$, plus the sum of the values
of all the self-energy graphs containing also self-energies
graphs with height $k+1$, which are absent in the
self-energy graphs contributing to $M^{[k]}(\oo\cdot\nn;\e)$.
Such a result will be used in Appendix \secc(A7) in order
to prove the following result.

\*

\sub(4.3) {\cs Lemma.}
{\it The power series defining the functions $\ohh^{[k]}_{\nn}(\e)$,
truncated at order $k'\le k$, coincide
with the functions $ \hh^{(\le k')}_{\nn}(\e)$ given by \equ(4.2).}

\*\*
\section(5,Convergence of the renormalized perturbative expansion)

\sub(5.1) {\it Domains of analyticity and norms.}
Consider in the complex $\e$-plane the domain $D_{\e_{0}}(\f)$
in Figure 5 below: if $\f$ denotes the half-opening of the sector
$D_{\e_{0}}(\f)$, then the radius of the circle delimiting
$D_{\e_{0}}(\f)$ will be of the form $\e(\f)=(\p-\f) \e_{0}$ (see below).

We shall define $\|\cdot\|$ an algebraic matrix norm
(\ie a norm which verifies $\| AB \| \le \| A \| \, \| B \|$
for all matrices $A$ and $B$);
for instance $\|\cdot\|$ can be the uniform norm.

The propagators $\ol{G}^{[k]}_{\ell}$ in \equ(4.4) satisfy interesting
$k$-independent bounds described and proved below.

\midinsert
\*
\eqfig{100pt}{90pt}
{\ins{72pt}{84pt}{$\e$-plane}
\ins{46pt}{58pt}{$D_{\e_{0}}(\f)$} \ins{9pt}{37pt}{$-x^{2}$}
} {fig5}{}
\*\*
\line{\vtop{\line{\hskip2.5truecm\vbox{\advance\hsize by -5.1 truecm
\0{{\css Fig. 5.}
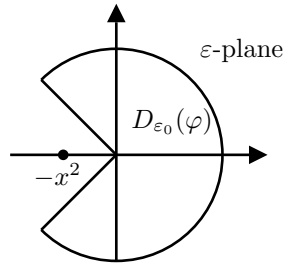
{\nota The domain $\st D_{\e_{0}}(\f)$ in the complex
$\st \e$-plane: the half-opening angle of the sector is $\st \f < \p$
but, otherwise, arbitrary, and the radius of the circle delimiting
$D_{\e_{0}}(\f)$ is given by $\e(\f)=(\p-\f)\e_{0}$.\vfill}}
} \hfill} }}
\*
\endinsert

\sub(5.2) {\cs Proposition.} {\it 
Let $D_{\e_{0}}(\f)$ be obtained from the disk of diameter $\e_{0}>0$
in the complex $\e$-plane by taking out a sector of half-opening $\p-\f$
around the negative real axis. Assume that the propagators
$\ol{G}^{[k]}_{\ell}\=\ol{G}^{[k]}(\oo\cdot\nn_{\ell};\e)$ satisfy 
$$ \left( \ol{G}^{[k]}(x;\e) \right)^{T} = \ol{G}^{[k]} (-x;\e),\qquad
\big\| \ol{G}^{[k]}(x;\e) \big\|
< \fra2{\p-\f}\fra1{x^2}
\Eq(5.1) $$
for all $|\e|<(\p-\f)\e_{0}$, if $\e_0$ is small enough.
Then there is a constant $B_{f}$ such that, summing over
all renormalized trees $\th$ with $|V(\th)|=V$ nodes the values
$|\oV^{[k]}(\th)|$, one has
$$ \eqalign{
& \ohh^{[k,V]}_{\g\nn} \le
\sum_{\th\in \Th^{\RR}_{V,\nn,\g}} \left| \oV^{[k]}(\th) \right|
\le  \left( \fra{|\e| B_{f}}{\p-\f} \right)^{V}, \cr
& \sum_{\th\in \Th^{\RR}_{V,\nn,\g} \atop M(\th) = s}
\left| \oV^{[k]}(\th) \right| \le
\left( \fra{|\e| B_{f}}{\p-\f} \right)^{V} e^{-\k s/8} , \cr
& \left| \bbb^{[k,V]}_{\V0} \right| \le
\left\| \left( \dpr_{\bb}^{2}f_{\V0}(\bb_{0})^{-1} \right) \right\|
\left( \fra{B_{f}}{\p-\f} \right)^{V+1} |\e|^{V}, \cr}
\Eq(5.2) $$
for all $s>0$.}

\*

\sub(5.3) {\it Remark.} 
Note that, although the propagators are no longer diagonal,
they still satisfy the same property as \equ(2.15), which is
the crucial one which is used in order to prove
both the lemmata \secc(2.5) and \secc(2.7) about the
formal solubility of the equations of motion and
the lemmata \secc(3.9) and \secc(3.11)
about the formal cancellations between tree values.

\*

\sub(5.4) {\it Proof of the proposition \secc(5.2).}
We can consider first trees without leaves, so that the tree
values are given by \equ(4.3) with $L(\th)=\emptyset$.

The hypothesis \equ(5.1) implies
that for all propagators $\ol{G}^{[k]}_{\ell}$ one has
$$ \left\| \ol{G}^{[k]}_{\ell} \right\| \le
C_{1} 2^{-2n_{\ell}} , \qquad C_{1} =
{2 \over \p - \f} \left( {2^{\t+2} \over C_{0} } \right)^{2} .
\Eq(5.3) $$
Therefore the contribution from a single tree (see \sec(2.4))
is bounded for all $n_{0}\le 0$ by 
$$ |\e|^{V} \left( C_{1}\,2^{-2(n_0-1)} \right)^{V} 
\prod_{v\in V(\th)}\Big[
\fra1{p_v!q_v!} |\nn_v|^{p_v+1}\dpr_{\bb}^{q_v+1}f_{\nn_v}(\bb_0)
\Big( \prod_{n=-\io}^{n_{0}-1} C_{1} 2^{-2 c |\nn_v| n 2^{n/\t}}
\Big) \Big] ,
\Eq(5.4)$$
where $V=|V(\th)|$, having used that, for all trees
$\th\in\Th_{k,\nn,\g}^{\RR}$, the number $N_{n}(\th)$ of lines
with scale $n$ in $\th$ satisfy the bound
$$ N_{n}(\th)  \le c\, M(\th) \, 2^{n/\t} =
c \, 2^{n/\t}\, \sum_{v\in V(\th)} |\nn_{v}| ,
\Eq(5.5) $$
for some constant $c$: an estimate which follows
from the proof of the lemma \secc(3.6) (see \sec(A3.4)).
The bound \equ(5.5) is used in deriving \equ(5.3) for all lines
$\ell\in\L(\th)$ with scale $n_{\ell}< n_0$, while for the lines
$\ell$ with scale $n_{\ell} \ge n_0$ we have used simply that the
propagators are bounded by $C_{1}2^{-2(n_{0}-1)}$.

If we use (recall that we are supposing that there are no leaves)
$$ \eqalign{
& {1 \over p!} |\nn|^{p+1}
\le (p+1) \left( {8 \over \k } \right)^{p+1} e^{\k |\nn|/8} , \cr
& {1 \over q!} \left| \dpr_{\bb}^{q+1} f_{\nn}(\bb_{0}) \right| \le
C_{2}^{q} F e^{-\k|\nn|} , \cr
& \sum_{v\in V(\th)} \left( p_{v} + q_{v} \right)
= k -1 , \cr}
\Eq(5.6) $$
for some constant $C_{2}$, and if we choose $n_{0}$ so that
$$ {\k \over 8} + 2 c \sum_{n=|n_0|+1}^{\io} n 2^{-n/\t}
\le {\k \over 4} ,
\Eq(5.7) $$
(\eg we can choose $n_{0}=\min\{0, -2\t\log2
\log((1-2^{-1/\t})\k/(16 c \log 2))$, then
we obtain the first bound in \equ(5.2), where we can take
$$ B_{f} = D_{0} C_{0}^{2} 2^{-(n_0-1)} F
\sum_{\nn\in\zzz^{r}} e^{-\k|\nn|/4} ,
\Eq(5.8)$$
for some positive constant $D_{0}$.
This follows after summing over all renormalized trees with
$V$ nodes and without leaves: this can be esaily done.
The sum over the mode labels can be performed by using
the decay factors $e^{-\k |\nn_{v}|/8}$, while the sum over all
the possible tree shapes gives a constant to the power $k$.

Furthermore the value $\k/8$ is so small that with our choices of the
constants an extra factor $\exp[-\k M(\th)/8]$
has been bounded by $1$ so that if, instead, the value of
$M(\th)$ is fixed we obtain the second bound of \equ(5.2).

So far we considered only trees without leaves.
If we want to consider also tree with leaves,
we can proceed in the following way.

Given a tree $\th$ (with leaves) of order $k'$,
we can write its renormalized value $\oV^{[k]}(\th)$ as the
product of the value of a trimmed tree $\ol{\th}$ times the
factors of its leaves: simply look at \equ(4.3)),
and interpret the renormalized value of the trimmed tree $\ol{\th}$ as
$$ \oV^{[k]}(\ol{\th}) =
\Big( \prod_{v\in V(\th)} F_{v} \Big)
\Big( \prod_{\ell \in \L(\th)} \ol{G}^{[k-1]}_{\ell} \Big) ,
\Eq(5.9) $$
while the product
$$ \Big( \prod_{v\in L(\th)} L_{v} \Big)
\Eq(5.10) $$
represents the product of the leaf factors \equ(4.6)
associated to the $|L(\th)|$ leaves of $\th$;
note that in \equ(5.9) we can completely neglect the
propagators associated to the lines exiting from the leaves,
as \equ(2.13) trivially implies.

The only effect of the leaves on $\oV^{[k]}(\ol{\th})$ is
through the presence of some extra derivatives $\dpr_{\bb}$ acting
on the node factors corresponding to some nodes $v\in V(\ol{\th})$;
in particular the momenta of the lines $\ell\in \L(\ol{\th})$
are completely independent of the leaves
(which contribute $\V0$ to such momenta).

Each leaf whose factor contributes to \equ(5.10) can be written as
sum of values of renormalized trees $\th_{1},\ldots,\th_{|L(\th)|}$,
according to \equ(4.6); for each of such tree, say $\th_{j}$,
we can write $\oV^{[k]}(\th_{j})$
as product of the renormalized value of the trimmed tree $\ol{\th}_{j}$
times the product the factors of its $|L(\th_{j})|$ leaves.
And so on: we iterate until only trimmed trees are left.
The sum of the orders of all trimmed trees equals the
order $k'$ of the tree $\th$.

Then we can see how the analysis performed above
in the case of trees without leaves can be modified
when also trees with leaves are taken into account.

First of all note that if, when considering the trees
whose renormalized values contribute to the leaf factor
\equ(4.6), we retain only the trees without leaves,
we can repeat the analysis leading to \equ(5.8),
with the only difference that (as it can be read from \equ(4.6))
one has a matrix $\left[ \dpr_{\bb}^{2}f_{\V0}(\bb_{0}) \right]^{-1}$
acting on the reduced value ${\oV^{[k]}}'(\th)$
and the tree $\th$ has order $\k_{v}+1$ (hence $V+1$,
if $V$ is the number of nodes of $\th$, as we are
supposing that $\th$ has no leaves), so that the
first bound in \equ(5.2) has to be replaced with
$$ \eqalign{
\left| \bbb^{[k,V]}_{\V0} \right| & \le \Big|
- \left[ \dpr_{\bb}^{2}f_{\V0}(\bb_{0}) \right]^{-1}
\sum_{\th\in \Th^{\RR*}_{V,\nn,\g}} {\oV^{[k]}}'(\th) \Big| \cr
& \qquad \le  \left\| \left( \dpr_{\bb}^{2}f_{\V0}(\bb_{0})^{-1}
\right) \right\| \left( \fra{D_{f}}{\p-\f} \right)^{V+1} |\e|^{V}, \cr}
\Eq(5.11) $$
so that we have an extra factor
$$ C_{3} = \left\| \left( \dpr_{\bb}^{2}f_{\V0}(\bb_{0})^{-1}
\right) \right\| \left( \fra{D_{f}}{\p-\f} \right)
\Eq(5.12) $$
with respect to the bound one obtains for \equ(5.2): this yields
the third bound in \equ(5.2) for leaves arising from
trees which do not contain other leaves.

Now we consider any tree of order $k'$, and we decompose
it in a collection of trimmed trees (as described above)
$\ol{\th}_{0}$, $\ol{\th}_{1}$, $\ol{\th}_{1}$, $\ldots$,
such that the root of $\ol{\th}_{0}$ is the root $r$ of $\th$,
while the root $r_{i}$ of each other trimmed tree $\ol{\th}_{i}$,
$i\ge 1$, coincides with a node of some other trimmed tree.
Moreover the propagators of the root lines of the  
trimmed trees $\ol{\th}_{j}$, $j\ge 0$,
can be neglected by the definition \equ(2.13).
Then the value of the tree $\th$ becomes
the product of (factorising) values of trimmed trees.

Then we can define the clusters as done in Section \secc(3),
with the further constraint that all lines interanl to a cluster
have to belong to the same trimmed tree.
Then for each trimmed tree the cancellation mechanisms
described in the previous Sections apply, and for each
of them the same bound as before is obtained.

Therefore for $\ol{\th}_{0}$ we can repeat the same analysis
as for trees without leaves with the only difference that
the third of \equ(5.6) does not hold anymore, and it
has to be replaced with
$$ \sum_{v\in V(\th)} \left( p_{v} + q_{v} \right) =
k-1 + |L(\th)| ;
\Eq(5.13) $$
as noted before the presence of the leaves implies
that, for each of them, there is a
derivative $\dpr_{\bb}$ acting
on the node factor of some node $v\in V(\ol{\th})$,
so that, with respect to the bound \equ(5.2),
we obtain an extra factor $C_{2}^{|L(\th)|}$ (one for leaf).

Now we can consider the trimmed trees $\ol{\th}_{1},\ldots,
\ol{\th}_{|L(\th)|}$, and proceed in the same way.
With respect to the previous case, for each trimmed
tree $\ol{\th}_{j}$ we obtain an extra factor $C_{2}$
for each leaf attached to some node of $\ol{\th}_{j}$.
Furthermore, as all the trimmed trees except $\ol{\th}_{0}$
contribute to leaves, there is also an extra factor
$C_{3}$ for each of them.

At the end, instead of the first bound in \equ(5.2) with
$D_{f}$ given by \equ(5.8), we obtain
$$ \left( {|\e|B_{f} \over \p-\f} \right)^{k'}
\left( C_{2} C_{3} \right)^{L} ;
\Eq(5.14) $$
as the total number of leaves is less than the total
number of lines with vanishing momentum (hence less than $k'$),
we obtain the first bound in \equ(5.2), provided that one
replaces the previous value for $B_{f}$ with
$$ B_{f} = D_{f} C_{2} C_{3} .
\Eq(5.15) $$
The sum over the trees can be performed exactly as in
the previous case.

In the same way one discusses the second and the
third bound in \equ(5.2), which follow with the constant
$B_{f}$ given by \equ(5.15). This completes the proof. \qed

\*

\sub(5.5) {\cs Proposition.} {\it Let $D_{\e_{0}}(\f)$ be as in the
proposition \secc(5.2); then the matrices $M^{[k]}(\oo\cdot\nn;\e)$
satisfy for $\e\in D_{\e_{0}}(\f)$ the relation
$$ \left( M^{[k]}(x;\e) \right)^{T} = M^{[k]}(-x;\e) .
\Eq(5.16) $$
Let also
$$ M^{[k]}(\oo\cdot\nn;\e) = \left( \matrix{
M^{[k]}_{\a\a}(\oo\cdot\nn;\e) & M^{[k]}_{\a\b}(\oo\cdot\nn;\e) \cr
M^{[k]}_{\b\a}(\oo\cdot\nn;\e) & M^{[k]}_{\b\b}(\oo\cdot\nn;\e) \cr}
\right) ;
\Eq(5.17) $$
then, if $\g_{q-1}<|x|\le \g_q$ and $\e_{0}$ is small enough,
the submatrices $M^{[k]}_{\g\g'}(x;\e)$ can be analytically continued
in the full disk $|x|\le \g_q$ and satisfy the bounds
$$ \eqalign{
& \left\| M^{[k]}_{\a\a}(x;\e) \right\| \le
\left( |\e|/(\p-\f) \right)^2\,C\,  x^{2} , \cr
& \left\| M^ {(k)}_{\a\b}(x;\e) \right\| \le
\left( |\e|/(\p-\f) \right)^2\,C\,  \left| x \right| , \cr
& \left\| M^{[k]}_{\b\b}(x;\e)- \e\,\dpr_{\bb}^{2} f_{\V0}(\bb_0)
\right\| \le \left( |\e|/(\p-\f) \right)^2\,C\, x^{2}  , \cr}
\Eq(5.18) $$
for all $k\in\NNN$ and for a suitable constant $C$.
As a consequence $\ol{G}^{[k]}_{\ell}$ verify\equ(5.1) for
all $k \ge 1$, and therefore \equ(5.2) holds for all $k \ge 1$.}

\*

\sub(5.6) {\cs Proof of the Proposition \secc(5.5).} We consider the
matrices $M^{[k]}$ defined in \equ(4.5) and suppose
{\it inductively} that $M^{[k]}$ verifies \equ(5.18)
and the analyticity property preceding it for $0\le k\le p-1$;
note that the assumption holds trivially for $k=0$.
Note also that \equ(5.18) imply that the propagators
$\ol{G}^{[k]}_{\ell}$ verify \equ(5.1) for $\e_{0}$ small enough
and $\e\in D_{\e_{0}}(\f)$.

To define $M^{[k]}$ we must consider the renormalized self-energy
graphs $T$ and evaluate their values by using the propagators
$\ol{G}^{[p-1]}_{\ell}$, according to \equ(4.5).

Given $x=\oo\cdot\nn$ such that $\g_{q-1} \le |x| < \g_{q}$ for
some $q\le 0$, the propagators $\ol{G}^{[p-1]}_{\ell}$
have an analytic extension to the disk $|x|<\g_{q+2}$ and,
under the hypotheses \equ(5.16) and \equ(5.18),
verify the symmetry property and the bound in \equ(5.1),
as it shown in Appendix \secc(A8).

We have (see \equ(4.5)) 
$$ M^{[p]}(x;\e) = \sum_{h = q+3}^{0}
\sum_{{\rm renormalized} \; T \atop n_{T} = h } \VV^{[p]}_{T,h}(x) ,
\Eq(5.19) $$
where by appending the label $h$ to $\VV^{[p]}_{T}(x)$
we distinguish the contributions to $M^{[p]}(x;\e)$
coming from self-energy graphs $T$ on scale $h$
(which is constrained to be $\ge q+3$; see the remark \secc(3.12), (2)).

The value $\VV^{[p]}_{T}(x)$  is analytic in $x$ for $|x|\le \g_{h+2}$
and the sum over all $T$'s with $V$ nodes is bounded by
$$ \sum_{T \atop |V(T)|=V} \left|\VV^{[p]}_{T,h}(x) \right|
\le \fra{(|\e| B_f)^V}{1-e^{-\k/8}} e^{-\k 2^{-h/\t}/8} ,
\Eq(5.20) $$
because the mode labels $\nn_{v}$ of the nodes $v\in V(T)$
must satisfy $\sum_{v\in V(T)} |\nn_v| > 2^{-h/\t}$
(recall that we are dealing with renormalized trees,
so that for all clusters $T\in \TT(\th)$ one has $T_{0}=T$,
and use \equ(3.8) and use the remark \secc(3.12), (2)).

Since the symmetry property expressed by \equ(5.1) for $k=p$
is implied by \equ(5.6) and this is the only property of
the propagators that one needs in order to check the algebraic
lemmata \secc(3.9) and \secc(3.11) (see the remark \secc(5.3)),
we can conclude that the same cancellation mechanisms
extend to the renormalized self-energy values
$\VV^{[p]}_{T}(\oo\cdot\nn)$ (see the remarks
\secc(A4.6) and \secc(A5.3)). Therefore
we see that $\VV^{[p]}_{T,h,\g\g'}(x)$ will vanish
at $x=0$ to order $\s_{\g\g'}$, if we set
$$ \s_{\g\g'} = \cases{
2, & for $\g=\a$ and $\g'=\a$, \cr
1, & for $\g=\a$ and $\g'=\b$, \cr
1, & for $\g=\b$ and $\g'=\a$, \cr
0, & for $\g=\b$ and $\g'=\b$; \cr}
\Eq(5.21) $$
moreover $\VV^{[p]}_{T,h,\b\b}(x)-\VV^{[p]}_{T,h,\b\b}(0)$
vanishes to order 2 at $x=0$.

By the analyticity in $x$ for $|x|\le \g_{h+2}$ and by the maximum
principle (Scharz's lemma) we deduce from \equ(5.10) that one has
$$ \eqalign{
& \sum_{T \atop |V(T)|=V} \left|\VV^{[p]}_{T,h,\g\g'}(x) \right|
\le \fra{(|\e| B_{f})^V}{1-e^{-\k/8}} e^{-\k 2^{-h/\t}/8}
\left(\fra{x}{\g_{h+2}}\right)^{\s_{\g\g'}} , \cr
& \sum_{T \atop |V(T)|=V} \left| \VV^{[p]}_{T,h,\b\b}(x) -
\VV^{[p]}_{T,h,\b\b}(0) \right|
\le \fra{(|\e| B_{f})^V}{1-e^{-\k/8}} e^{-\k 2^{-h/\t}/8}
\left(\fra{x}{\g_{h+2}}\right)^2 , \cr}
\Eq(5.22)$$
Therefore we can use that $\sum_{h=q+3}^{0} e^{-\k 2^{-h/\t}}
2^{-2h}< B_1<\io$ and that $V\ge 2$ for
$(\g,\g') \in \{(\a,\a),(\a,\b),(\b,\a)\}$, while
$V\ge1$ for $(\g,\g')=(\b,\b')$, and the proof is complete. \qed

\*

\sub(5.7) {\it Convergence of the sequence
$\{M^{[k]}(\oo\cdot\nn;\e)\}_{k\in\NNN}$.} 
It also follows that there exists the limit
$$ \lim_{k\to\io} M^{[k]}(x;\e) = M^{[\io]}(x;\e) ,
\Eq(5.23) $$
with $M^{[\io]}(x;\e)$ analytic in $\e$ in $D_{\e_{0}}(\f)$: in fact the
following result holds (the proof is in Appendix \secc(A8)).

\*

\sub(5.8) {\cs Lemma.} {\it For all $k\ge 1$ one has
$$ \left\| M^{[k+1]}(x;\e) - M^{[k+1]}(x;\e) \right\|
\le \hat B_{1} \hat B_{2}^{k} \e_{0}^{2k} ,
\Eq(5.24)$$
for some constants $\hat B_{1}$ and $\hat B_{2}$.}

\*

\sub(5.9) {\it Fully renormalized expansion.}
We can now define the ``fully renormalized'' expansion of the
parametric equations of the invariant torus as the sum of the
values of the renormalized trees evaluated according
to \equ(4.3) with $\ol{G}_{\ell}^{[k-1]}$ replaced by 
$$ \ol{G}^{[\io]}(x;\e) = \left( x^{2} \openone-
M^{[\io]}(x;\e) \right)^{-1} , \qquad
x = \oo\cdot\nn_{\ell} .
\Eq(5.25) $$
The above discussion shows that the series converges for all
$\e\in D_{\e_{0}}$ and that it coincides
with the limit for $k\to\io$ of $\ohh^{[k]}(\pps;\e)$,
which therefore exists.

The radius of the domain $D_{\e_{0}}(\f)$ is $\left(\p-\f\right)\e_{0}$,
if $\f$ is the half-opening of the sector $D_{\e_{0}}(\f)$,
because the norms of the propagators $\ol{G}^{[\io]}(x;\e)$
are bounded by $2/(x^{2}(\p-\f))$ (see \secc(A8.3)).

Therefore the functions $\ohh^{[k]}(\pps;\e)$ converge in a
heart-like domain
$$ \bigcup_{-\p \le \a < \p} D_{\e_{0}}(\f) = D_{0} ,
\Eq(5.26) $$
whose boundary, for negative $\e$ close to $0$,
is such that $\Im(\e)$ is proportional to $(\Re(\e))^{2}$.

\*

\sub(5.10) {\cs Proposition.} {\it There exist positive constants
$\e_{0}$, $B$, $\tilde B_{1}$ and $\tilde B_{2}$, such that if
$$ \eqalign{
&\hh^{(k)}_{\RR\g\nn}(\e) =
\sum_{\Th^{\RR}_{k,\nn,\g} } \oV^{[\io]}(\th) , \cr
& \oV^{[\io]}(\th) =
\Big( \prod_{v\in V(\th)} F_{v} \Big)
\Big( \prod_{v\in \L(\th)} F_{v} \Big)
\Big( \prod_{\ell \in L(\th)} \ol{G}_{\ell}^{[\io]} \Big) , \cr}
\Eq(5.27)$$
the renormalized series 
$$ \ohh^{[\io]}(\pps;\e) =
\sum_{k=1}^{\io} \e^{k} \sum_{\nn\in\zzz^{r}} e^{i\nn\cdot\pps}
\hh^{(k)}_{\RR\nn}(\e) 
\Eq(5.28) $$
converges in the heart-shaped domain \equ(5.26) and its
coefficients are bounded by
$$ \left| \hh^{(k)}_{\RR\nn}(\e) \right|
\le \tilde B_{1} \tilde B_{2} ^{k} , \qquad
N!^{-1}\left| \dpr_\e^N \hh^{(k)}_{\RR\nn}(\e) \right| <
N!^{2\t+1} B^N\,
\tilde B_{1} \tilde B_{2}^{k} , \qquad {\rm for}\; N\ge0 ,
\Eq(5.29) $$
uniformly in $\e\in D_{0}$.}

\*

\sub(5.11) {\it Comments about \equ(5.29).}
We leave out, for simplicity, the proof that the $N!^{2\t+1}$ is the
appropriate power of $N!$ that follows from our analysis. Although it
is quite clear that one has obtained a remainder bound proportional to
a power of $N!$, we evaluated it explicitly in the hope that the power
series expansion of $\ohh^{[\io]}(\pps;\e)$
(hence of $\hh(\pps;\e)$, see the proposition \secc(5.13) below)
at $\e=0$ could be shown to be summable in
the sense of the Borel transforms or of its extensions. Since we have
analyticity of $\V h$ in the domain $D_{0}$ of Figure 1 in Section
\secc(1) we would need that the remainder in \equ(5.29)
behaves at most as $N!^2$, see \cita{CGM}.
Since $\t\ge r-1$ and $r\ge2$ (in order to have quasi-periodic
solutions) we see that \equ(5.29) is not
compatible with the general theory. Therefore one needs more
information than just \equ(5.29) in order to be able to reconstruct
from the power series at the origin the full equation of the invariant
torus.

\*

\sub(5.12) {\it Conclusions.} 
In Appendix \secc(A10) we show that the function
$\ohh^{[\io]}(\pps;\e)$, \ie the limit for $k\to\io$ of the
approximated functions $\ohh^{[k]}(\pps;\e)$, solves the equations of
motion \equ(1.6), so proving the following proposition: this concludes
the proof of the theorem \secc(1.3).
\*

\sub(5.13) {\cs Proposition.}
{\it One has, formally (\ie order by order in the expansion in $\e$
around $\e=0$)
$$ \ohh^{[\io]}(\pps;\e) \=
\lim_{k\to\io} \ohh^{[k]}(\pps;\e) = \hh(\pps;\e) ,
\Eq(5.30) $$
where $\hh(\pps;\e)$ is the formal power series which solves
the equations \equ(1.6).}

\*\*
\section(6,Concluding remarks)

\sub(6.1) {\it Some extensions.} The case 
of more general Hamitonians of the form
$$ \HHH= h_{0}(\AA) + \e f(\aa,\AA) ,
\Eq(6.1) $$
with $(\aa,\AA)\in\TTT^{d}\times\AAA$, where
$\AAA$ is an open domain in $\RRR^{d}$, should be easily studied as
the case treated here to show existence and regularity of invariant
tori associated with rotation vectors $\oo\in\RRR^{d}$
among whose components there are $s$ rational relations,
while the independent ones verify a Diophantine condition.

\*

\sub(6.2) {\it Periodic orbits.} The fully resonant case $r=1$
corresponds to periodic orbits is of course a special case of
our theory, but it is well known. Note that in such a case the
series expansion envisaged in Section \secc(2) is sufficient to prove
existence (and analyticity) of the periodic solutions,
and no resummation is needed; see the remark \secc(2.10).

\*

\sub(6.3) {\cs Conjecture.} 
{\it An obvious conjecture is that the above functions can be further
continued analytically and that the continuation touches the real
negative axis of the $\e$-plane in a set of points with $0$ as a
Lebesgue density point. We hope to be able to prove this result that
seems within the reach of the technique introduced here.}

\*

\sub(6.4) {\it (Lack of) Borel summability.}
As pointed out in the concluding sentence of Section \secc(5)
the results that we have are not sufficient to imply
(extended) Borel summability of the formal power series
at the origin of the parametric equations of
the torus, \ie of $\hh(\pps;\e)$. The resummations that lead to the
construction of $\hh(\pps;\e)$ are therefore of a different type from
the well known ones associated with the Borel transforms.

\*\*
\0{\bf Acknowledgements.} We are indebted to V. Mastropietro for his
criticism in the early stages of this work.
We thank also H. Eliasson for useful discussions.

\*\*
\appendix(A1,Proof of the lemma {\secc(2.5)})

\asub(A1.1) {\it Proof of the lemma. Part I: neglecting the factorials.}
Consider all contributions arising from
the trees $\th\in\Th_{\V0,k,\a}$: we group together
all trees obtained from each other by shifting the root line,
\ie by changing the node which the root line exits
and orienting the arrows in such a way that they still
point toward the root.
We call $\FF(\th)$ such a class of trees (here $\th$
is any element inside the class).

The reduced values ${\Val}'(\th')$ of such trees
$\th'\in\FF(\th)$ differ because

\0(1) there is a factor $i\nn_{v}$ depending on
the node $v$ to which the root line is attached
(see the definition \equ(2.9) of $F_{v}$), and

\0(2) some arrows change their directions;
more precisely, when the root line is detached from the node $v_{0}$
and reattached to the node $v$, if $\PP(v_{0},v)=\{ w\in V(\th)
\;:\; v_{0} \succeq w \succeq v\}$ denotes the
path joining the node $v_{0}$ to the node $v$,
all the momenta flowing through the lines $\ell$
along the path $\PP(v_{0},v)$ change their signs,
the factorials of the node factors corresponding to the nodes
joined by them can change, and
the propagators $G_{\ell}$ are replaced with their transposed.

The change of the signs of the momenta simply
follows from the fact that
$$ \sum_{v\in V(\th)} \nn_{v} = \V0 ,
\Eqa(A1.1) $$
as $\th\in\Th_{k,\V0,\a}$: by the property \equ(2.15),
the propagator does not change.

The change of the factorials contributing to the node factors
is due to the fact that for the nodes along the path $\PP(v_{0},v)$,
an entering line can become an exiting line and {\it vice versa},
so that the labels $p_{v}$ and $q_{v}$ can be transformed
into $p_{v}\pm 1$ and $q_{v}\pm 1$, respectively:
this does not modify the factor $(i\nn_{v})^{p_{v}+(1-\d_{v})}
\dpr_{\bb}^{q_{v}+\d_{v}}f_{\nn_{v}}(\bb_{0})$
in \equ(2.9) -- up to the factor $i\nn_{v}$ (if the root line
is attached to $v$), which has been
already taken into account --, as one immediately checks,
but it can produce a change of the factorials.

If we neglect the change of the factorials, \ie
if we assume that all combinatorial factors are the same,
by summing the reduced values of all possible trees inside the class
$\FF(\th)$ we obtain a common value times $i$ times
\equ(A1.1), and the sum gives zero.

\*

\asub(A1.2) {\it Proof of the lemma. Part II:
taking into account the factorials.}
One can easily show that a correct counting of the trees
implies that all factorials are in fact equal:
to do this it is convenient to use {\it topological} trees
instead of the usual {\it semitopological} used so far
(we follow the discussion in \cita{BeG}).

We briefly outline the differences between the two kinds
of trees, deferring to \cita{G2} and \cita{GM}
for a more detailed discussion of the differences
between what finally amounts to a different way to count trees.
Define a group of transformations acting on trees generated
by the following operations: fix any node $v\in V(\th)$ and permute
the subtrees entering such a node. We shall call semitopological
trees the trees which are superposable up to a continuous
deformation of the lines, and topological trees the trees
for which the same happens modulo the action of the just
defined group of transformations.
We define {\it equivalent} two trees
which are equal as topological trees.

Then we can still write \equ(2.15) restricting the sum
over the set of all nonequivalent
topological trees of order $k$ with labels $\nn_{\ell_{0}}=\nn$
and $\g_{\ell_{0}}=\g$ (we can denote it by
$\Th_{k,\nn,\g}^{\rm top}$), provided that to each node
$v \in V(\th)$ we associate a combinatorial factor which is
not the $(p_{v}!q_{v}!)^{-1}$ appearing in \equ(2.9).

In fact for topological trees the
combinatorial factor associated to each node is different, because we
have to	look now to how	the subtrees emerging from each	node differ. 
For semitopological trees we have a factor $(p_{v}!q_{v}!)^{-1}$
for each node $v$, where $p_{v}$ and $q_{v}$ are the numbers of
lines $\ell$ with $\g_{\ell}=a$ and $\g_{\ell}=\b$, respectively,
entering $v$: except for the labels $\g_{\ell}'$,
we are disregarding the kinds of the subtrees entering $v$, so that
in this way we are counting as different many trees
otherwise identical. On the contrary, in the
case of	topological trees, we consider one and the same tree those
trees that are different as semitopological trees, but have the	same
value because they just	differ in the order in which {\it identical}
subtrees enter each node $v$: therefore, if $s_{v,1}$, \dots,
$s_{v,j_{v}}$ are the number of	entering lines to which	are
attached subtrees of a given shape and with the same
labels (so that $s_{v,1}+\dots+s_{v,j_{v}}=p_{v}+q_{v}$,
$1 \le j_{v} \le p_{v}+q_{v}$), the combinatorial factor,
for each node, becomes
$$ {1 \over p_{v}!q_{v}!} \cdot {p_{v}!q_{v}! \over
s_{v,1}! \ldots s_{v,j_{v}}!} = 
{1 \over s_{v,1}! \ldots s_{v,j_{v}}! } ;
\Eqa(A1.2) $$
note in	the second factor in the above formula the multinomial 
coefficient corresponding to the number	of different 
semitopological trees corresponding to the same 
topological tree, for each node.

So in terms of topological trees $\aaa^{(k)}_{\nn}$ and
$\bbb^{(k)}_{\nn}$ can be expressed as sum of tree values
${\Val}^{\rm top}(\th)$, where
$$ {\Val}^{\rm top}(\th) = 
\Big( \prod_{v\in V(\th)} F_{v}^{\rm top} \Big)
\Big( \prod_{v\in L(\th)} L_{v} \Big)
\Big( \prod_{\ell \in \L(\th)} G_{\ell} \Big) ,
\Eqa(A1.3) $$
where
$$ F_{v}^{\rm top} = {1\over s_{v,1}! \ldots s_{v,j_{v}}! }
\Big( i\nn_{v} \Big)^{p_{v}+(1-\d_{v})}
\dpr_{\bb}^{q_{v}+\d_{v}} f_{\nn_v}(\bb_{0}) .
\Eqa(A1.4) $$

Still, when computing the combinatorial factors inside each
family $\FF(\th)$, they do differ.
But this is actually an apparent, not a real discrepancy.
In fact, due to symmetries in the tree
(that is, to the fact that the subtrees emerging from some node
are sometimes equal, \ie that some $s_{v,i}$ are greater than $1$),
the actual number of topological trees in a given family
$\FF(\th)$ is less than the total number of trees
obtained by the action of the group of transformations:
in other words some trees obtained by the action of the group
are equivalent as topological trees.
When moving the	root line from a node $v_{0}$ to another
node $v_{1}$, so transforming a	tree $\th$ into	a tree $\th_{1}\in
\FF(\th)$, for some nodes $w$ along the path $ P(v_{0},v_{1})$
the factor $1/s_{w,i}!$ can turn into $1/(s_{w,i}-1)!$, but then
this means that the same topological tree could be formed by the
action of $s_{w,i}$ different transformations of the group:
each of the $s_{w,i}$ equivalent subtrees entering $w$
contains a node such that, by attaching to it
the root line, the same	topological tree is obtained.
Therefore, by counting all trees obtained by the
action of the group,
the corresponding topological tree value is in fact counted
$s_{w,i}$ times, so to avoid overcounting one needs a factor
$1/s_{w,i}$: this gives back the same combinatorial factor $1/s_{w,i}!$.
Analogously one	discusses the case of a	factor $1/s_{w,i}!$
turning	into $1/(s_{w,i}+1)!$, simply by noting that
the same argument as above can be followed also in this case
by changing the	r\^oles	of the two nodes $v_{0}$ and $v_{1}$.

\*

\asub(A1.3) {\it Remark.} The proof of the lemma relies
only on the property \equ(2.15) of the propagators,
so that also the function $\ohh^{[k]}(\pps;\e)$
is well defined for all $k\in\NNN$.

\*\*
\appendix(A2,Proof of the lemma {\secc(2.7)})

\asub(A2.1) {\it Proof.}
In order to prove the lemma we shall show by induction
that $\aaa^{(k)}_{\V0}$ can be arbitrarily fixed
and $\bbb^{(k)}_{\V0}$ can be uniquely fixed
in order to make formally solvable the equations \equ(2.2).

For $k=1$ it is straightforward to realize that $\aaa^{(1)}_{\nn}$
and $\bbb^{(1)}_{\nn}$ are well defined for all
$\nn \in \ZZZ^{r}\setminus\{\V0\}$,
by using the first condition in \equ(1.8).

Then, for $k>1$, assume that all $\aaa^{(k')}_{\V0}$
and $\bbb^{(k')}_{\V0}$, with $k'<k-1$, have been fixed,
and that, as a consequence, all $\aaa^{(k')}_{\nn}$ and
$\bbb^{(k')}_{\nn}$ are well defined for $k'<k$ and for all
$\nn\in\ZZZ^{r}\setminus\{\V0\}$.

By \equ(2.16) and by the lemma \secc(2.7), in \equ(2.2)
one has $\left[\dpr_{\aa}f \right]^{(k-1)}_{\nn}=0$,
so that the equation for $\aaa^{(k)}$ is
formally soluble, and $\aaa^{(k)}_{\V0}$
can be arbitrarily fixed, for instance equal to $\V0$.

In the second equation in \equ(2.2) one can write
$$ \left[ \dpr_{\bb} f \right]^{(k)}_{\nn} =
\dpr_{\bb}^{2} f_{\V0}(\bb_{0}) \bbb^{(k-1)}_{\nn} + \GG_{\nn}^{(k)} ,
\Eqa(A2.1) $$
where the function $\GG_{\nn}^{(k)}$ takes into account
all contributions except the one explicitly written,
and, by construction, all terms appearing in
$\GG_{\nn}^{(k)}$ can depend only on factors
$\bbb^{(k')}_{\V0}$ of orders $k' \leq k-2$. We can choose
$$ \bbb^{(k-1)}_{\V0} = - \Big[ \dpr_{\bb}^{2} f_{\V0}(\bb_{0})
\Big]^{-1} \GG_{\V0}^{(k)} ,
\Eqa(A2.2) $$
where the second condition in \equ(1.8) has been used,
so that also the equation for $\bbb^{(k)}$ becomes
formally soluble. Of course also $\bbb^{(k)}_{\V0}$
is left undetermined: it will have to be fixed in the
next iterative step.

To complete the proof of the lemma one has still to show
that the sums over the Fourier labels can be performed,
but this is a trivial fact for $\oo\in D_{\t}(C_{0})$.

\*

\asub(A2.2) {\it Remark.} The same proof applies to the
renormalized trees introduced in \sec(4.1).

\*\*
\appendix(A3,Proof of the lemma {\secc(3.6)})

\asub(A3.1) {\it Inductive bounds.} 
We prove inductively on the number of nodes of the trees the bounds
$$
N_{n}^{*}(\th)
\le \max\{ 0 , 2\, M(\th) \, 2^{(n+3)/\t} - 1 \} ,
\Eqa(A3.1) $$
where $M(\th)$ is defined in \equ(3.15).

First of all note that if $M(\th)<2^{-(n+3)/\t}$ then
$N_{n}(\th)=0$ as in such a case for any line $\ell\in\L(\th)$ one has
$$ \left| \oo_{0}\cdot\nn_{\ell} \right| >
2^{\t} |\nn_{\ell}|^{-\t} > 2^{\t} M(\th)^{-\t} > 2^{\t}2^{n+3} ,
\Eqa(A3.2) $$
by the Diophantine hypothesis \equ(1.2) and by the definition
of $\oo_{0}$ given in \sec(3.2).

\*

\asub(A3.2) {\it Bound on $N_{n}^{*}(\th)$.}
If $\th$ has only one node the bound is trivially satisfied because,
if $v$ is the only node in $V(\th)$,
one must have $M(\th)=|\nn_{v}|\ge 2^{-n/\t}$
in order that the line exiting from $v$ is on
scale $\le n$: then $2\,M(\th)\,2^{(n+3)/\t}\ge 4$.

If $\th$ is a tree with $V>1$ nodes, we assume that
the bound holds for all trees having $V'<V$ nodes.
Define $E_{n}=(2\,2^{(n+3)/\t})^{-1}$: so we have
to prove that $N_{n}^{*}(\th)\le \max\{0,M(\th)\,E_{n}^{-1}-1\}$.

If the root line $\ell$ of $\th$ is either on scale $\neq n$ or a
self-energy line with scale $n$, call $\th_{1},\ldots,\th_{m}$
the $m\ge 1$ subtrees entering the last node $v_{0}$ of $\th$.
Then
$$ N_{n}^{*}(\th) = \sum_{i=1}^{m} N_{n}^{*}(\th_{i}) ,
\Eqa(A3.3) $$
hence the bound follows by the inductive hypothesis.

If the root line $\ell$ is normal (\ie it is not
a self-energy line) and it has scale $n$, call
$\ell_{1},\ldots,\ell_{m}$ the $m\ge 0$ lines on scale $\le n$ which are
the nearest to $\ell$ (this means that no other line along the paths
connecting the lines $\ell_{1},\ldots,\ell_{m}$ to the
root line is on scale $\le n$). Note that in such a case
$\ell_{1},\ldots,\ell_{m}$ are the entering line
of a cluster $T$ on scale $>n$.

If $\th_{i}$ is the subtree with $\ell_{i}$ as root line, one has
$$ N_{n}^{*}(\th) = 1 + \sum_{i=1}^{m} N_{n}^{*}(\th_{i}) ,
\Eqa(A3.4) $$
so that the bound becomes trivial if either $m=0$ or $m\ge 2$.

If $m=1$ then one has $T=\th\setminus\th_{1}$, and the
lines $\ell$ and $\ell_{1}$ are both with scales $\le n$;
as $\ell_{1}$ is not entering a self-energy graph, then
$$ |\oo_{0}\cdot\nn_{\ell}|\le 2^{n} , \qquad
|\oo_{0}\cdot\nn_{\ell_{1}}|\le 2^{n} ,
\Eqa(A3.5) $$
and either $\nn_{\ell} = \nn_{\ell_{1}}$ and one must have
(recall that $T_{0}$ is defined after \equ(3.6))
$$ \sum_{v\in V(T)} |\nn_{v}| \ge \sum_{v\in V(T_{0})} |\nn_{v}| > \
2^{-(n+3)/\t} = 2E_{n} > E_{n} ,
\Eqa(A3.6) $$
or $\nn_{\ell}\neq\nn_{\ell_{1}}$, otherwise $T$ would be a
self-energy graph (see \equ(3.7) and \equ(3.8)).
If $\nn_{\ell}\neq\nn_{\ell_{1}}$, then, by \equ(A3.5)
one has $|\oo_{0}\cdot ( \nn_{\ell}-\nn_{\ell_{1}})|
\le 2^{n+1}$, which, by the Diophantine condition \equ(1.2),
implies $|\nn_{\ell}-\nn_{\ell_{1}}|>2\,2^{-(n+1)/\t}$,
so that again
$$ \sum_{v\in V(T)} |\nn_{v}| \ge
\left| \nn_{\ell}-\nn_{\ell_{1}} \right|
> 2 \, 2^{-(n+2)/\t} > 2^{1/\t+2} E_{n} > E_{n} ,
\Eqa(A3.7) $$
as in \equ(A3.6).
Therefore in both cases we get
$$ M(\th)-M(\th_{1}) = \sum_{v\in T} |\nn_{v}| > E_{n} ,
\Eqa(A3.8) $$
which, inserted into \equ(A3.4) with $m=1$, gives,
by using the inductive hypothesis,
$$ \eqalign{
N_{n}^{*}(\th) & = 1 + N_{n}^{*}(\th_{1})
\le 1 + M(\th_{1})\,E_{n}^{-1} - 1 \cr
& \le 1 + \Big( M(\th) - E_{n}\Big) E_{n}^{-1} - 1
\le M(\th) \, E_{n}^{-1} - 1 , \cr}
\Eqa(A3.9) $$
hence the bound is proved also if the root line
is normal and on scale $n$.

\*

\asub(A3.4) {\it Remark.} The same argument proves the
bound (5.4) for renormalized trees, by using the observation
that there are no self-energy lines in the renormalized trees.

\*\*
\appendix(A4,Proof of the lemma {\secc(3.9)})

\asub(A4.1) {\it Factorials.}
As for the proof of the lemma \secc(2.7) we ignore the factorials:
to take them into account one can reason as said in Appendix \secc(A1).

\*

\asub(A4.2) {\it Self-energy graphs of type 1.}
First we prove that $\VV_{T}(0)=0$.
Given a tree $\th$ consider all trees which can be obtained
by shifting the entering line $\ell_{T}^{2}$.
Note that the trees so obtained are contained in the
self-energy graph family $\FF_{T_{0}}(\th)$.

Corresponding to such an operation $\VV_{T}(0)$ changes
by a factor $i\nn_{v}$ if $v$ is the node which the
entering line is attached to, as all node factors
and propagators do not change.
By \equ(3.6) the sum of all such values is zero.

Then consider $\dpr\VV_{T}(0)$. By construction
$$ \dpr\VV_{T}(0) = \sum_{\ell\in \L(T) }
\Big( \prod_{v\in V(T) } F_{v} \Big)
\Big( \dpr G^{(n_{\ell})}_{\ell}
\prod_{\ell'\in \L(T) \setminus\ell}
G^{(n_{\ell'})}_{\ell'} \Big) ,
\Eqa(A4.1) $$
where all propagators have to be computed for $\oo\cdot\nn=0$, and
$$ \dpr G^{(n_{\ell})}_{\ell} = \Big.
{\der \over \der x} G^{(n_{\ell})}_{\ell}
(\oo\cdot\nn_{\ell}^{0} + \s_{\ell} x ) \Big|_{x=0} , \qquad
x = \oo\cdot\nn .
\Eqa(A4.2) $$
The line $\ell$ divides $V(T)$ into two disjoint set of
nodes $V_{1}$ and $V_{2}$, such that
$\ell_{T}^{1}$ exits from a node inside $V_{1}$ and
$\ell_{T}^{2}$ enters a node inside $V_{2}$:
if $\ell=\ell_{v}$ one has $V_{2}=\{w\in \V(T) : w \preceq v\}$
and $V_{1}=V(T)\setminus V_{2}$. By \equ(3.4), if
$$ \nn_{1} = \sum_{v\in V_{1}} \nn_{v} ,
\qquad \nn_{2} = \sum_{v\in V_{2}} \nn_{v} ,
\Eqa(A4.3) $$
one has $\nn_{1}+\nn_{2}=\V0$.
Then consider the families $\FF_{1}(\th)$ and $\FF_{2}(\th)$
of trees obtained as follows: $\FF_{1}(\th)$
is obtained from $\th$
by detaching $\ell_{T}^{1}$ then reattaching to all the nodes
$w\in V_{1}$ and
by detaching $\ell_{T}^{2}$ then reattaching to all the nodes
$w\in V_{2}$, while $\FF_{2}(\th)$
is obtained from $\th$
by reattaching the line $\ell_{T}^{1}$ to all the nodes
$w\in V_{2}$ and
by reattaching the line $\ell_{T}^{2}$ to all the nodes
$w\in V_{2}$;
note that $\FF_{1}(\th)\cup\FF_{2}(\th) \subset \FF_{T_{0}}(\th)$.

As a consequence of such an operation the arrows
of some lines $\ell\in\L(T)$
change their directions: this means that for some line $\ell$
the momentum $\nn_{\ell}$ is replaced with $-\nn_{\ell}$
and the propagators $G_{\ell}$
are replaced with their transposed $G^{T}_{\ell}$.
As the propagators satisfy \equ(2.15)
no overall change is produced by such factors,
except for the differentiated propagator which can change sign:
one has a different sign for the trees in $\FF_{1}(\th)$
with respect to the trees in $\FF_{2}(\th)$.
Then by summing over all the possible trees in $\FF_{1}(\th)$
we obtain a value $i^{2}\nn_{1}\nn_{2}$ times a common factor,
while by summing over
all the possible trees in $\FF_{2}(\th)$
we obtain $-i^{2}\nn_{1}\nn_{2}$ times the same common factor,
so that the sum of two sums gives zero.

\*

\asub(A4.3) {\it Self-energy graphs of type 2.}
Given a tree $\th$
with a self-energy graph $T$ consider all trees obtained
by detaching the exiting line, then reattaching
to all the nodes $v\in V(T)$;
note again that the trees so obtained are contained in the
self-energy graph family $\FF_{T_{0}}(\th)$.
In such a case again some momenta can change sign,
but the corresponding propagator does not change
(reason as above for self-energy graphs of type 1).
So at the end we obtain a common
factor times $i\nn_{v}$, where $v$ is the node which
the exiting line is attached to. By \equ(3.6) again
we obtain $\VV_{T}(0)=0$.

\*

\asub(A4.4) {\it Self-energy graphs of type 3.}
To prove that $\VV_{T}(0)=0$ simply reason as for $\VV_{T}(0)$
in the case (1), by using that the entering line
$\ell_{T}^{2}$ has $\g_{\ell_{T}^{2}}=\a$.

\*

\asub(A4.5) {\it  Self-energy graphs of type 4.}
Given a tree $\th$
with a self-energy graph $T$ consider the contribution to
$\dpr V_{T}(0)$ in which a line $\ell$ is differentiated (see \equ(A4.1)).
The line $\ell$ divides $V(T)$ into two disjoint set of
nodes $V_{1}$ and $V_{2}$, such that
$\ell_{T}^{1}$ exits from a node $v_{1}$ inside $V_{1}$ and
$\ell_{T}^{2}$ enters a node $v_{2}$ inside $V_{2}$:
if $\ell=\ell_{v}$ one has $V_{2}=\{w\in V(T) : w \preceq v\}$
and $V_{1}=V(T)\setminus V_{2}$. By \equ(3.6), with
the notations \equ(A4.3), one has $\nn_{1}+\nn_{2}=\V0$.
Then consider the tree obtained
by detaching $\ell_{T}^{1}$ from $v_{1}$,
then reattaching to the node $v_{2}$ and, simultaneously,
by detaching $\ell_{T}^{2}$ from $v_{2}$,
then reattaching to the node $v_{1}$; note that the tree
so obtained is inside the class $\FF_{T_{0}}(\th)$.

As a consequence of such an operation the arrows
along the path $\PP$ connecting $v_{1}$ to $v_{2}$
change their directions: this means that for such lines $\ell$
the momentum $\nn_{\ell}$ is replaced with $-\nn_{\ell}$,
but the propagators are even in the momentum,
so that no overall change is produced by such factors,
if not because of the differentiated propagator (which
is along the path by construction) which changes sign.
For all the other lines (\ie the lines not belonging to $\PP$)
the propagator is left unchanged.

Since a derivative with respect to $\bb$ acts on both the nodes
$v_{1}$ and $v_{2}$, the shift of the external lines
does not produce any change on the node factors
(except for the factorials, that we are not explicitly
considering, as said at the beginning of this subsection).
Then by summing over the two considered trees we obtain zero
because of the change of sign of the differentiated propagator.

\*

\asub(A4.6) {\it Remark.}
To prove the lemma \secc(3.9) we only use that the
propagators satisfy \equ(2.15), so that the same
proof applies also to the renormalized self-energy graphs
(see the proposition \secc(5.5)), where there are no self-energy
lines and the propagators are given by \equ(4.4).

\*\*
\appendix(A5,Proof of the lemma {\secc(3.11)})

\asub(A5.1) {\it Proof of the property (1).}
Given a self-energy graph $T$ with momentum $\nn$
flowing through the entering line $\ell_{T}^{2}$,
call $\PP$ the path connecting the exiting line $\ell_{T}^{1}$
to the entering line $\ell_{T}^{2}$.
Then consider also the self-energy graph $T'$
obtained by taking $\ell_{T}^{1}$ as entering line and
$\ell_{T}^{2}$ as exiting line and
by taking $-\nn$ as momentum flowing through the (new)
entering line $\ell_{T}^{1}$:
in this way the arrows of all the lines along the path $\PP$
are reverted, while all the subtrees
(internal to $T$) having the root in $\PP$ are left unchanged.
This implies that the momenta of the lines belonging to $\PP$
change signs, while all the other momenta do not change.
Since all propagators $G_{\ell}$ are transformed
into $G_{\ell}^{T}$ the property \equ(2.15)
implies that the entry $ij$ of the matrix $M(\oo\cdot\nn;\e)$
corresponding to the self-energy graph $T$ is equal to the entry
$ji$ of the matrix $M(-\oo\cdot\nn;\e)$; then the assertion follows.

\*

\asub(A5.2) {\it Proof of the property (2).}
Given a self-energy graph $T$,
consider also the self-energy graph $T'$
obtained by reverting the sign of the mode labels
of the nodes $v\in V(T)$, and
by swapping the entering line with the exiting one.
In this way the arrows of all the lines along the path $\PP$
joining the two external lines are reverted, while all the subtrees
(internal to $T$) having the root in $\PP$ are left unchanged.
It is then easy to realize that the complex conjugate
of $\VV_{T'}(\oo\cdot\nn)$ equals $\VV_{T}(\oo\cdot\nn)$,
by using the form of the node factors \equ(2.10), and the fact
that one has $f_{\nn}^{*}(\bb)=f_{-\nn}(\bb)$
and $G^{\dagger}(\oo\cdot\nn)=G(\oo\cdot\nn)$ (see \equ(2.15)).

\*

\asub(A5.3) {\it Remark.} The lemma has been proved
without making use of the exact form of the propagator, but only
exploiting the fact that it satisfies the property \equ(2.15):
therefore, once more, the proof applies also to the renormalized trees
as a consequence of the first relation in \equ(5.1),
and it gives \equ(5.6).

\*\*
\appendix(A6,Proof of the lemma {\secc(3.14)})

\asub(A6.1) {\it Set-up.}
Given a tree $\th\in\Th_{k,\nn,\g}$, consider
a self-energy graph $T$ with height $D$.
Call $T^{(2)} \subset T^{(3)} \subset \ldots \subset T^{(D)}$
the resonances containing $T$, and set $T=T^{(1)}$;
denote by $n=n_{1}>n_{2}>n_{3}>\ldots>n_{D}$ the scales
of the lines entering such resonances.

For any $\ell\in \L(T_{0})$, one can write
$$ \nn_{\ell}=\nn_{\ell}^{0}+\s_{\ell}\nn ,
\Eqa(A6.1) $$
where $\nn$ is the momentum of the line $\ell_{T}^{2}$
(with scale $n$) entering $T$ (see \equ(3.9)).

\*

\asub(A6.2) {\it Proof.}
By shifting the lines entering all the resonances
containing $\ell$, the momentum $\nn_{\ell}$ can change into a new
value $\tilde\nn_{\ell}$ (as it can be seen
by applying iteratively \equ(3.10)) in such a way that
$\left|\oo_{0}\cdot\tilde\nn_{\ell}\right|$
differs from $\left|\oo_{0}\cdot\nn_{\ell}^{0}\right|$
by a quantity bounded by $\g_{n}+\g_{n_{2}} + \ldots
+ \g_{n_{D}}$ $\le$ $2^{n}+2^{n_{2}}+\ldots+2^{n_{D}}< 2^{n+1}$.

As $\left|\nn_{\ell}^{0}\right|<2^{-(n+3)/\t}$,
by definition of self-energy graph, we can apply \equ(3.2)
and conclude that $|\oo_{0}\cdot\nn_{\ell}^{0}|$
has to be contained inside an interval $[\g_{p-1},\g_{p}]$,
with $p=n_{\ell}^{0} \ge n+3$ (see the remark \secc(3.12), (2)),
at a distance at least $2^{n+1}$ from the extremes:
therefore the quantity $\left|\oo_{0}\cdot\tilde\nn_{\ell}\right|$
still falls inside the same interval $[\g_{p-1},\g_{p}]$.
In particular this implies the identity
$$ n_{\ell} = \tilde n_{\ell} = n_{\ell}^{0} ,
\Eqa(A6.4) $$
if $\tilde n_{\ell}$ is defined as the integer such that
$\g_{\tilde n_{\ell}-1} \le
\left|\oo_{0}\cdot\tilde\nn_{\ell}\right| < \g_{\tilde n_{\ell}}$.

\*\*
\appendix(A7,Proof of the lemma {\secc(4.3)})

\asub(A7.1) {\it Set-up.} We have to prove that, for all $k'\le k$,
one has
$$ [ \ohh^{[k]}(\pps;\e) ]^{(k')}_{\nn} = \hh^{(k')}_{\nn} ,
\Eqa(A7.1) $$
which yields that the functions $\hh$ and $\ohh^{(\le k)}$ admit the
same power series expansions up to order $k$.

Of course both $\hh^{(k')}_{\nn}$ and
$[\ohh^{[k]}(\pps;\e)]^{(k')}_{\nn}$ are given by sums
of several contributions; in the same way
$[M^{[k]}(\oo\cdot\nn;\e)]^{(k')}$
can be expressed as sum of several terms.
We shall prove that, given any tree value $\Val(\th)$
contributing to $\hh^{(k')}_{\nn}$ and
any self-energy value $\VV_{T}(x)$
corresponding to a self-energy graph $T$ of order $k'$,
one can find the same terms contributing, respectively,
to $[\ohh^{[k]}(\pps;\e)]^{(k')}_{\nn}$ and
to $[M^{[k]}(x;\e)]^{(k')}$, and {\it vice versa}.
The proof will be by induction on $k'=1,\ldots,k$.

\*

\asub(A7.2) {\it Remarks.} (1)
Note that $[ \ohh^{[k]}(\pps;\e)]^{(k')}_{\nn}$ is not the same
as $\ohh^{[k,k']}_{\nn}(\pps;\e)$: the quantity
$[ \ohh^{[k]}(\pps;\e)]^{(k')}_{\nn}$ is the
coefficient to order $k'$ that one obtains
by developping $\ohh^{(k)}(\pps;\e)$ in powers of $\e$.
\\
(2) The contributions to $[\ohh^{[k]}(\pps;\e)]^{(k')}_{\nn}$
and to $[M^{[k]}(x;\e)]^{(k')}$ can arise only
from trees of order $k''\le k'$. Furthermore
$[\ohh^{[k]}(\pps;\e)]^{(k')}_{\nn}=[\ohh^{[k]}(\pps;\e)]^{(k')}_{\nn}$
and $[M^{[k]}(x;\e)]^{(k')}=[M^{[k]}(x;\e)]^{(k')}$:
this symply follows from the remark \secc(4.2) and the
trivial observation that, for $k>k'$,
the self-energy graphs [trees] whose values contribute to
$M^{[k]}(x;\e)$ [$\ohh^{[k]}(\pps;\e)$]
but not to $M^{[k']}(x;\e)$ [$\ohh^{[k']}(\pps;\e)$]
are those containing also self-energy graphs with height $D>k'$:
such contributions are of order at least $D$ in $\e$,
so that they can not contribute to
$[M^{[k]}(x;\e)]^{(k')}$ [$[\ohh^{[k]}(\pps;\e)]^{(k')}$].

\*

\asub(A7.3) {\it Starting from $\hh$.}
The case $k'=1$ is trivial. Suppose that, given $k' \le k$,
the assertion is true for all $k''<k'$:
then we show that is is true also for $k'$.

Consider a tree $\th\in\Th_{k,\nn,\g}$ and let $\Val(\th)$
be its value. Denote by $T_{1},\ldots,T_{N}$ the self-energy graphs
in $\th$ with height $D=0$, and by $k_{1},\ldots,k_{N}$ the number
of nodes that they contain, respectively; the number of nodes
external to the self-energy graphs will be $k_{0}=k-k_{1}-\ldots-k_{N}$.
Call $\th_{0}$ the tree obtained from $\th$ by replacing
each chain of self-energy graphs together with their external lines
with a new line carrying the same momentum of the external lines.
The tree $\th_{0}$ will have $k_{0}$ lines: by construction
each line $\ell_{i}$ of $\th_{0}$ corresponds to a chain of
$p_{i}$ resonances, with $p_{i}\ge 0$, of orders
$K_{i1},\ldots,K_{ip_{i}}$, such that
$$ \sum_{i=1}^{k_{0}} \sum_{j=1}^{p_{i}} K_{ij} = k - k_{0} .
\Eqa(A7.2) $$

Then consider $\th_{0}$ as a tree $\th^{\RR}\in\Th^{\RR}_{k_{0},\nn,\g}$;
let $\ell_{i}$ the line in $\th^{\RR}$ which corresponds
to the line with the same name in $\th_{0}$.
For each line $\ell_{i}\in\L(\th^{\RR})$, by setting
$x_{i}=\oo\cdot\nn_{\ell_{i}}$, the propagator is of the form
$$ \ol{G}_{\ell_{i}}^{[k-1]} = \ol{G}^{[k-1]}(x_{i};\e) , \qquad
\ol{G}^{[k-1]}(x_{i};\e) = \left[ x_{i}^{2} \openone -
M^{[k-1]}(x_{i};\e) \right]^{-1} ,
\Eqa(A7.3) $$
and it can be expanded in powers of $M^{[k-1]}(x_{i};\e)$ as
$$ \ol{G}^{[k-1]}(x_{i};\e) = {1\over x_{i}^{2}} \left( \openone
+ M^{[k-1]}(x_{i};\e) {1\over x_{i}^{2}} +
M^{[k-1]}(x_{i};\e) {1\over x_{i}^{2}}
M^{[k-1]}(x_{i};\e) {1\over x_{i}^{2}} + \ldots \right) .
\Eqa(A7.4) $$
We can consider the contribution
$$ {1 \over x_{i}^{2}} [M^{[k-1]}(x_{i};\e)]^{(K_{i1})}
{1 \over x_{i}^{2}} \ldots 
{1 \over x_{i}^{2}} [M^{[k-1]}(x_{i};\e)]^{(K_{ip_{i}})}
{1 \over x^{2}}
\Eqa(A7.5) $$
to \equ(A7.4). Note that $K_{ij} < k-1$ for all $i,j$, by
construction, so that by the remark \secc(A7.2), (2), we can write
$$ [M^{[k-1]}(x_{i};\e)]^{(K_{ip_{i}})} =
[M^{[k]}(x_{i};\e)]^{(K_{ip_{i}})} =
[M^{[K_{ip_{i}}-1]}(x_{i};\e)]^{(K_{ip_{i}})} ,
\Eqa(A7.6) $$
for all $j=1,\ldots, p_{i}$ and for all $i=1,\ldots,k_{0}$.
Hence by the inductive hypothesis, we can decuce
that, for all $i,j$, there is a contribution
to $[M^{[k-1]}(x_{i};\e)]^{(K_{ij})}=
[M^{[k]}(x_{i};\e)]^{(K_{ij})}$ which corresponds
to the considered resonance in $\th$. As a consequence
we can also conclude that there is a term contributing to
$[\ohh^{[k]}(\pps;\e)]^{(k')}_{\nn}$
which is the same as the considered tree value $\Val(\th)$.

Of course if instead of a tree value we had considered
a self-energy value, the same argument should have applied,
so that the assertion follows.

\*

\asub(A7.4) {\it Starting from $\ohh^{[k]}$.}
The construction described in \secc(A7.3) can be 
used in the opposite direction, in order to prove
that each term of order $k'$ in $\epsilon$ which
is obtained by truncating $\ohh^{[k]}$ to order $k$
corresponds to a term contributing to $\hh^{(k')}$.

\*\*
\appendix(A8,Proof of the bound {\equ(5.1)} from {\equ(5.8)})

\asub(A8.1) {\it Set-up.} Consider the matrix
$$ A(x;\e) = \left( G^{[k]}(x;\e) \right)^{-1}
= x^{2}\openone - M^{[k]}(x;\e) =
\L + \e^{2}x\D_{1} + \e^{2}x^{2}\D_{2} ,
\Eqa(A8.1) $$
with
$$ \eqalign{
\L & = \left( \matrix{ \L_{\a\a} & 0 \cr 0 & \L_{\b\b} \cr} \right)
= \left( \matrix{ x^{2}\openone & 0 \cr 0 &
x^{2}\openone - \e \dpr^{2}_{\bb}f_{\V0}(\bb_{0}) \cr} \right) , \cr
\D_{1} & = \left( \matrix{ 0 & B(\e) \cr
-B(\e) & 0 \cr} \right) , \cr
\D_{2} & = \left( \matrix{
- M^{[k]}_{\a\a}(x;\e) & - M^{[k]}_{\a\b}(x;\e) - \e^{2}xB(\e) \cr
- M^{[k]}_{\b\a}(x;\e) - \e^{2}xB(\e) & -M^{[k]}_{\b\b}(x;\e)
+ \e\dpr^{2}_{\bb}f_{\V0}(\bb_{0}) \cr} \right) , \cr}
\Eqa(A8.2) $$
where
$$ B(\e) = \left( \e^{2} x \right)^{-1}
\sum_{T'} \LL \VV_{T'}(\oo\cdot\nn) ,
\Eqa(A8.3) $$
with the sum running over all self-energy graphs of type 2,
so that one has
$$ \left\| \D_{1} \right\| \le C , \qquad
\left\| \D_{2} \right\| \le C ,
\Eqa(A8.4) $$
for some positive constant $C$.

The matrix $\L$ is a block matrix which induces  natural
decomposition $\RRR^{d}=\RRR^{r}\oplus \RRR^{s}$;
the eigenvalues of the block $\L_{\a\a}\=\L|\RRR^{r}$ are all equal
to $x^{2}$, while the eigenvalues of the block $\L_{\b\b}=\L|\RRR^{s}$
are of the form $\l_{j}=x^{2}+a_{j}\e$, with $a_{j}>0$.

Set $\L_{1}=\L+\h \D_{1}$, with $\h=\e^{2}x$.

Define
$$ \eqalign{
B(x;\e) & = e^{\h X}A(x;\e)e^{-\h X} =
e^{\h X}\L_{1}e^{-\h X} + \e^{2}x^{2} e^{\h X}\D_{2}e^{-\h X}
\cr & \= B_{0}(x;\e) + \e^{2}x^{2} e^{\h X}\D_{2}e^{-\h X} ; \cr}
\Eqa(A8.5) $$
of course $B(x;\e)$ has the same eigenvalues as $A(x;\e)$.

\*

\asub(A8.2) {\it Block-diagonalization.}
Consider $B_{0}(x;\e)$: we shall fix the matrix $X$ in such a way
that $B_{0}(x;\e)$ is block-diagonal up to order $\h^{2}$.

So we look for $X$ such that
$$ \left( \openone + \h X + O(\h^{2}) \right)
\left( \L + \h \D_{1} \right)
\left( \openone - \h X + O(\h^{2}) \right)
= \L + \h J_{1} + O(\h^{2}) ,
\Eqa(A8.6) $$
with
$$ J_{1} = \left( \matrix{
J_{1,\a\a} & J_{1,\a\b} \cr J_{1,\b\a} & J_{1,\b\b} \cr} \right) =
\left( \matrix{
J_{1,\a\a} & 0 \cr 0 & J_{1,\b\b} \cr} \right) .
\Eqa(A8.7) $$
By expanding to first order \equ(A8.6) we obtain
$$ \left[ X,\L \right] + \D_{1} = J_{1} ,
\Eqa(A8.8) $$
while imposing \equ(A8.7) gives
$$ \eqalign{
X_{\a\b} & = - B(\e) \left( \L_{\b\b} - x^{2}\openone \right)^{-1} , \cr
X_{\b\a} & = - \left( \L_{\b\b} - x^{2}\openone \right)^{-1} B(\e) , \cr}
\Eqa(A8.9) $$
where
$$ \left\| \left( \L_{\b\b}-x^{2} \openone \right)^{-1} \right\|
= {1 \over \e} \left\| \left( \dpr_{\bb}^{2} f_{\V0} (\bb_{0})
\right)^{-1} \right\| \le {C \over \e} ,
\Eqa(A8.10) $$
for some constant $C$.
Furthermore, by choosing $X_{\a\a}=0$ and $X_{\b\b}=0$,
one obtains $J_{1}\=0$.

Then it follows that one has
$$ B(x;\e) = \L + O(\h^{2}X^{2}) + O(\e^{2}x^{2}) =
\L + O(\e^{2}x^{2}) ,
\Eqa(A8.11) $$
so that
$$ B^{-1}(x;\e) = \L^{-1} + O(\e^{2}x^{2}) ,
\Eqa(A8.12) $$
where the eigenvalues of $\L^{-1}$ are of the form
either $1/x^{2}$ or $1/(x^{2}+a_{j}\e)$.

Therefore one has
$$ \left\| G^{[k]}(x;\e) \right\| \le
\left\| e^{\h X} \right\|
\left\| B^{-1} \right\|
\left\| e^{-\h X} \right\| \le {2 \over x^{2} } ,
\Eqa(A8.13) $$
which proves the lemma \secc(5.4).

\*

\asub(A8.3) {\it Bounds in the complex plane.}
The above analysis applies also for complex values of $\e$.
Consider the domain $D_{0}$ represented in Fig. 5,
with half-opening angle $\f<\p$, and set $\a=\p-\f$.
For $\e\in D_{0}$ one has that the norms of
$\ol{G}^{[k]}(x;\e))^{-1}$ are bounded from below by
$$ {x^{2} \over 2} \left( \p - \a \right) ,
\Eqa(A8.14) $$
when $\a$ is close to $0$.
\*

\*\*
\appendix(A9,Proof of the lemma {\secc(5.8)})

\asub(A9.1) {\it Set-up.} Both $M^{[k]}(\oo\cdot\nn;\e)$
and $M^{[k+1]}(\oo\cdot\nn;\e)$ can be expressed
by \equ(4.5): the only difference is that one has to use
the propagators $\ol{G}^{[k]}_{\ell}$ for
$M^{[k+1]}(\oo\cdot\nn;\e)$.

This means that there is a correspondence 1-to-1
between the graphs contributing to $M^{[k]}(\oo\cdot\nn;\e)$ and those
contributing to $M^{[k+1]}(\oo\cdot\nn;\e)$, so that we can write
$$ \eqalign{
& M^{[k]}(\oo\cdot\nn;\e) - M^{[k+1]}(\oo\cdot\nn;\e) = 
\sum_{{\rm renormalized}\;T} \VV^{[k,k+1]}_{T}(\oo\cdot\nn) , \cr
& \VV^{[k,k+1]}_{T}(\oo\cdot\nn) = 
\e^{|V(T)|} \, \Big( \prod_{v\in V(T)} F_{v} \Big)
\Big[ \Big( \prod_{\ell\in \L(T)} \ol{G}^{[k-1]}_{\ell} \Big)
- \Big( \prod_{\ell\in \L(T)} \ol{G}^{[k]}_{\ell} \Big) \Big] . \cr}
\Eqa(A9.1) $$
For each renormalized self-energy $T$
we can write $\VV^{[k,k+1]}_{T}(\oo\cdot\nn)$
as sum of $V=|V(T)|$ terms corresponding to trees
whose lines have all the propagators of the form either
$\ol{G}^{[k-1]}_{\ell}$ or $\ol{G}^{[k]}_{\ell}$,
up to one which has a new propagator given by the difference
$\ol{G}^{[k-1]}_{\ell}-\ol{G}^{[k]}_{\ell}$.

\*

\asub(A9.2) {\it Remark.} Note that
the scales of all lines are uniquely fixed by
the momenta, so that both propagators
$\ol{G}^{[k-1]}_{\ell}$ and $\ol{G}^{[k]}_{\ell}$
admit the same bounds (see \equ(5.6)).

\*

\asub(A9.3) {\it Bounds.} 
We can order the lines in $\L(T)$ and
construct a set of $V$ subsets
$\L_{1}(T),\ldots,\L_{V}(T)$ of $\L(T)$,
with $|\L_{j}(T)|=j$, in the following way.
Set $\L_{1}(T)=\emptyset$, $\L_{2}(T)=\ell_{1}$,
if $\ell_{1}$ is the root line of $\th$
and, inductively for $2\le j\le V-1$,
$\L_{j+1}(T)=\L_{j}(T)\cup\ell_{j}$,
where the line $\ell_{j}\in \L(T)\setminus \L_{j}(T)$ is
connected to $\L_{j}(T)$; of course $\L_{V}(T)=\L(T)$. Then
$$ \eqalign{
& \VV^{[k,k+1]}_{T}(\oo\cdot\nn) = 
\e^{|V(T)|} \Big( \prod_{v\in V(T)} F_{v} \Big)
\cr & \qquad
\sum_{j=1}^{V}
\Big[ \Big( \prod_{\ell\in \L_{j}(T)} \ol{G}^{[k-1]}_{\ell} \Big)
\left( \ol{G}^{[k-1]}_{\ell_{j}} - \ol{G}^{[k]}_{\ell_{j}} \right)
\Big( \prod_{\ell\in \L(T) \setminus \L_{j}(T)}
\ol{G}^{[k]}_{\ell} \Big) \Big] , \cr}
\Eqa(A9.2) $$
where, by construction, the sets $\L_{j}(\th)$ are connected
(while of course the sets $\L(\th)\setminus \L_{j}(\th)$
in general are not).

We can write
$$ \ol{G}^{[k-1]}_{\ell_{j}} - \ol{G}^{[k]}_{\ell_{j}} =
\ol{G}^{[k-1]}_{\ell_{j}} \left[
M^{[k]}(\oo\cdot\nn_{\ell_{j}};\e) -
M^{[k-1]}(\oo\cdot\nn_{\ell_{j}};\e) \right] \ol{G}^{[k]}_{\ell_{j}} ,
\Eqa(A9.3) $$
so that in \equ(A9.2) we can bound
$$ \eqalign{
& \Big( \prod_{\ell\in \L_{j}(T)}
\left\| \ol{G}^{[k-1]}_{\ell} \right\| \Big)
\left\| \ol{G}^{[k-1]}_{\ell_{j}} \right\|
\left\| \ol{G}^{[k]}_{\ell_{j}} \right\|
\Big( \prod_{\ell\in \L(T) \setminus \L_{j}(T)}
\left\| \ol{G}^{[k]}_{\ell} \right\| \Big) 
\cr
& \qquad \le \left[
\left( 2^{2+2\t}C_{0}^{-1} \right)^{2k} \prod_{n=-\io}^{1}
2^{-2nN_{n}(\th)} \right]^{2} , \cr}
\Eqa(A9.4) $$
where the power 2 (with respect to \equ(5.9))
is due to the fact that in the product
two propagators correspond to the line $\ell_{j}$.

This means that $\VV^{[k,k+1]}_{T}(\oo\cdot\nn)$
admits the same bound as the square
of $\VV^{[k]}_{T}(\oo\cdot\nn)$ times the norm
$$ \left\| M^{[k]}(\oo\cdot\nn;\e) -
M^{[k-1]}(\oo\cdot\nn;\e) \right\| .
\Eqa(A9.5) $$
If we perform the sum over all self-energy graphs
in \equ(A9.1) and we use that the first non-trivial
terms corresponds to graphs with $V=2$ nodes, we obtain,
for $k\ge 1$,
$$ \left\| M^{[k+1]}(\oo\cdot\nn;\e) -
M^{[k]}(\oo\cdot\nn;\e) \right\| \le C \e^{2}
\left\| M^{[k]}(\oo\cdot\nn;\e) -
M^{[k-1]}(\oo\cdot\nn;\e) \right\| ,
\Eqa(A9.6) $$
for some constant $C$. Therefore the lemma follows.

\*\*
\appendix(A10,Proof of the proposition {\secc(5.13)})

\asub(A10.1) {\it Set-up.} Define
$$ \oV^{[\io]}(\th) =
\Big( \prod_{v\in V(\th)} F_{v} \Big)
\Big( \prod_{v\in \L(\th)} F_{v} \Big)
\Big( \prod_{\ell \in L(\th)} \ol{G}_{\ell}^{[\io]} \Big) ,
\Eqa(A10.1) $$
where the propagators $\ol{G}^{[\io]}_{\ell}=
\ol{G}^{[\io]}(\oo\cdot\nn_{\ell};\e)$ are defined in \equ(5.19);
we shall denote by $\ol{G}^{[\io]}$
the operator with kernel $\ol{G}^{[\io]}(\oo\cdot\nn;\e)$
in Fourier space.

Then one has
$$ \ohh^{[\io]}_{\g}(\pps;\e)
= \sum_{k=1}^{\io} \sum_{\nn\in\zzz^{r}}
\e^{k} e^{i\nn\cdot\pps} \sum_{\th\in\Th^{\RR}_{k,\nn,\g}}
\oV^{[\io]}(\th) ,
\Eqa(A10.2) $$
which we can represent, in a more compact notations, as
%
%
%
$$ \ohh^{[\io]}(\pps;\e) = \sum_{\th\in\Th^{\RR}}
\oV^{\RR}(\th;\pps;\e) ,
\Eqa(A10.3) $$
where $\Th^{\RR}$ is the set of all renormalized
trees, and, for $\th\in\Th^{\RR}_{k,\nn,\g}\subset\Th^{\RR}$,
we have defined
$$ \oV^{\RR}(\th;\pps;\e) =
\e^{k} e^{i\nn\cdot\pps} \oV^{[\io]}(\th) .
\Eqa(A10.4) $$

The function $\hh(\pps;\e)$ solving the equations of motion
\equ(1.6) is formally defined
as the solution of the functional equation
$$ \hh(\pps;\e) = G \dpr_{\pps} f \left( \pps + \hh(\pps;\e) \right) ,
\Eqa(A10.5) $$
where $G=(i\o\cdot\dpr)^{-2}=\ol{G}^{[0]}$ is the operator
with kernel $G(x)=x^{2}$.

We have the following result.

\*

\asub(A10.2) {\cs Lemma.}
{\it One has $G(x) \left( M^{[\io]}(x;\e) +
(G^{[\io]}(x;\e))^{-1} \right)=\openone$.}

\*

\asub(A10.3) {\it Proof.}
By definition one has $\ol{G}^{[\io]}(x;\e)=
(G^{-1}(x) - M^{[\io]}(x;\e) )^{-1}$, so that
$G^{-1}(x)=(\ol{G}^{[\io]}(x;\e))^{-1} + M^{[\io]}(x;\e)$;
then the assertion follows. \qed

\*

\asub(A10.4) {\it Conclusions.}
The following result shows that the
function $\ohh^{[\io]}(\pps;\e)$
formally solves the equation of motions \equ(1.6);
as the analysis of the previous sections
shows that the function $\ohh^{[\io]}(\pps;\e)$
is well defined and it is, order by order,
equal to the formal solution envisged in Section \secc(3),
we have proved the proposition.

\*

\asub(A10.5) {\cs Lemma.} {\it The function $\ohh^{[\io]}(\pps;\e)$
defined by \equ(A10.3) formally solves \equ(A10.4).}

\*
 
\asub(A10.6) {\it Proof.}
We shall show that \equ(A10.3) solves \equ(A10.5). One has
%
%
$$ \eqalign{
& G \dpr_{\pps} f \left( \pps + \ohh^{[\io]}(\pps;\e) \right)
= G \sum_{p=0}^{\io} {1\over p!} \dpr_{\pps}^{p+1}
f(\pps) \left( \ohh^{[\io]}(\pps;\e) \right)^{p} \cr
& \qquad = G \sum_{p=0}^{\io} {1\over p!} \dpr_{\pps}^{p+1}
f(\pps) \sum_{\th_{1}\in\Th^{\RR}} \oV^{\RR}(\th_{1};\pps;\e)
\ldots \sum_{\th_{p}\in\Th^{\RR}} \oV^{\RR}(\th_{p};\pps;\e) \cr
& \qquad = G \left( \ol{G}^{[\io]} \right)^{-1}
\sum_{\th\in\Th_{\RR}^{*}} \oV^{\RR}(\th;\pps;\e) , \cr}
\Eqa(A10.7) $$
where $\Th_{\RR}^{*}$ differs from $\Th^{\RR}$ as it contains
also trees which can have only one self-energy graph
with exiting line $\ell_{0}$, if, as usual,
$\ell_{0}$ denotes the root line of $\th$; the operator
$G(\ol{G}^{[\io]})^{-1}$ takes into account the fact that,
by construction, to the root line $\ell_{0}$ an operator
$G$ is associated, while in $\oV^{\RR}(\th;\pps;\e)$,
by definition, a propagator $\ol{G}^{[\io]}$ is associated.

Then we can write \equ(A10.5), by explicitly separating
the trees containing such a self-energy graph from the others,
$$ \eqalign{
& G \dpr_{\pps} f \left( \pps + \ohh^{[\io]}(\pps;\e) \right) \cr
& \qquad = G \left( \ol{G}^{[\io]} \right)^{-1} \Big(
\ol{G}^{[\io]} M^{[\io]} \sum_{\th\in\Th^{\RR}} \oV^{\RR}(\th;\pps;\e)
+ \sum_{\th\in\Th^{\RR}} \oV^{\RR}(\th;\pps;\e) \Big) \cr
& \qquad = G \Big( M^{[\io]} \ohh^{[\io]}(\pps;\e)
+ (\ol{G}^{[\io]})^{-1} \ohh^{[\io]}(\pps;\e) \Big) \cr
& \qquad = G \Big( M^{[\io]} + (\ol{G}^{[\io]})^{-1} \Big)
\ohh^{[\io]}(\pps;\e) = \ohh^{[\io]}(\pps;\e) , \cr}
\Eqa(A10.8) $$
where the lemma \secc(A10.2) has been used in the last line.

Note that at each step only absolutely converging series
have been dealt with; then the assertion is proved.


\rife{BaG}{BaG}{M.V. Bartuccelli, G. Gentile:
{\it Lindstedt series for perturbations of isochronous systems.
General theory}, Rev. Math. Phys., in press. }
\rife{BeG}{BeG}{A. Berretti, G. Gentile:
{\it Scaling properties for the radius of convergence
of Lindstedt series: generalized standard maps},
J. Math. Pures Appl.
{\bf 79} (2000), no. 7, 691--713. }
\rife{BGGM}{BGGM}{F. Bonetto, G. Gallavotti, G. Gentile:
{\it Lindstedt series, ultraviolet divergences and Moser's theorem},
Ann. Scuola. Norm. Sup. Pisa Cl. Sci. (4)
{\bf 26} (1998), no. 3, 545--593. }
%
%
\rife{B}{B}{ A.D. Bryuno:
{\it Analytic form of differential equations. I, II. (Russian)},
Trudy Moskov. Mat. Ob\v{s}\v{c}.
{\bf 25} (1971), 119--262;
ibid. {\bf 26} (1972), 199--239. } 
\rife{BKS}{BKS}{J. Bricmont, A. Kupiainen, A. Schenkel:
{\it Renormalization group for the Melnikov problem for PDE's},
Comm. Math. Phys., in press. }
\rife{CGM}{CGM}{E. Caliceti, V. Grecchi, M. Maioli:
{\it The distributional Borel summability and the large
coupling $\Phi^4$ lattice fields},
Commun. Math. Phys. {\bf 104} (1986), 163--174. }
\rife{CG}{CG} {L. Chierchia, G. Gallavotti:
{\it Drift and diffusion in phase space},
Ann. Inst. H. Poincar\`e Phys. Th\'eor.
{\bf 60} (1994), no. 1, 1--144;
{\it Erratum for ``Drift and diffusion in phase
space'', by L. Chierchia and G. Gallavotti,
Ann. Inst. H. Poincar\'e Phys. Th\'eor.
60 (1994), 1, 1994},
Ann. Inst. H. Poincar\'e Phys. Th\'eor.
{\bf 68} (1998), 135. }
\rife{E}{E}{ Eliasson, L.H.:
{\it Absolutely convergent series expansions for quasi-periodic motions},
Math. Phys. Electron. J. {\bf 2} (1996), Paper 4, 
{\tt <http:// mpej.unige.ch>}. }
\rife{G1}{G1}{G. Gallavotti:
{\it Twistless KAM tori},
Comm. Math. Phys.
{\bf 164} (1994), 145--156. }
\rife{G2}{G2}{G. Gallavotti:
{\it Twistless KAM tori, quasiflat homoclinic intersections,
and other cancellations in the perturbation series of
certain completely integrable Hamiltonian systems. A review},
Rev. Math. Phys.
{\bf 6} (1994), no. 3, 343--411. }
\rife{GG}{GG}{G. Gallavotti, G. Gentile:
{\it Majorant series convergence for twistless KAM tori},
Ergodic Theory Dynam. Systems
{\bf 15} (1995), 857--869. }
\rife{Ge}{Ge}{G. Gentile:
{\it Whiskered tori with prefixed frequencies and Lyapunov spectrum},
Dynam. Stability of Systems
{\bf 10} (1995), no. 3, 269--308. }
\rife{GGM}{GGM}{G. Gentile, G. Gallavotti, V. Mastropietro:
{\it Field theory and KAM tori},
Math. Phys. Electron. J.
{\bf 1} (1995), Paper 1,
{\tt <http:// mpej.unige.ch>}. }
\rife{GM}{GM}{G. Gentile, V. Mastropietro:
{\it Methods for the analysis of the Lindstedt series for KAM tori and
renormalizability in Classical mechanics. A review with some applications},
Rev. Math. Phys.
{\bf 8} (1996), no. 3, 393--444. }
\rife{Gr}{Gr}{S.M. Graff:
{\it On the conservation for hyperbolic invariant
tori for Hamiltonian systems},
J. Differential Equations
{\bf 15} (1974), 1--69. }
\rife{Me}{Me}{V.K. Mel'nikov:
{\it On some cases of conservation of conditionally periodic motions
under a small change of the Hamiltonian function},
Soviet Math. Dokl.
{\bf 6 } (1965), 1592--1596;
{\it A family of conditionally periodic solutions of a
Hamiltonian systems},
Soviet Math. Dokl.
{\bf 9 } (1968), 882--886. }
\rife{Mo}{Mo}{J. Moser:
{\it Convergent series expansions for quasi periodic motions},
Math. Ann. {\bf 169} (1967), 136--176. }
\rife{P}{P} {J. P\"{o}schel:
Invariant manifolds of complex analytic mappings near fixed points,
in {\sl Critical Phenomena, Random Systems, Gauge Theories},
Les Houches, Session XLIII (1984), Vol. II, 949--964, Ed.
K. Osterwalder \& R. Stora, North Holland, Amsterdam, 1986. }
\rife{R}{R} {H. R\"ussmann:
{\it Invariant tori in the perturbation
theory of weakly non--degenerate integrable Hamiltonian systems}, 
Preprint-Rehie des Fachbereichs Mathematik, nr. 14, 1--89, 1998. }
\rife{XY}{XY} {J. Xu, J. You:
{\it Persistence of Lower Dimensional 
Tori Under the First Melnikov's Non-resonance Condition},
Nanjing University Preprint, 1--21, 2001. }

\biblio

\ciao